**Magnetic Properties from the Viewpoints of Electronic Hamiltonian: Spin Exchange Parameters, Spin Orientation and Spin-Half Misconception**

Myung-Hwan Whangbo[1,]* and Hongjun Xiang[2,3]

[1] Department of Chemistry, North Carolina State University, Raleigh, North Carolina 27695-8204, USA

[2] Key Laboratory of Computational Physical Sciences (Ministry of Education), State Key Laboratory of Surface Physics, and Department of Physics, Fudan University, Shanghai 200433, P. R. China

[3] Collaborative Innovation Center of Advanced Microstructures, Nanjing 210093, P. R. China

E-mails: mike_whangbo@ncsu.edu

**Abstract**

In this chapter we review the quantitative and qualitative aspects of describing the properties of magnetic solids on the basis of electronic Hamiltonian, which describes the energy states of a magnetic system using both orbital and spin degrees of freedom. To quantitatively discuss a magnetic property of a given magnetic system, one needs to generate the spectrum of its energy states and subsequently average the properties of these states with each state weighted by its Boltzmann distribution factor. Currently, this is an impossible task to achieve on the basis of an electronic Hamiltonian, so it is necessary to resort to a simple model Hamiltonian, i.e., a spin Hamiltonian that describes the energy states of a magnetic system using only the spin degree of freedom. We show that a spin Hamiltonian approach becomes consistent with an electronic Hamiltonian approach if the spin lattice and its associated spin exchange parameters, to be used for the spin Hamiltonian, are determined by the energy-mapping analysis based on DFT calculations. The preferred spin orientation (i.e., the magnetic anisotropy) of a magnetic ion is not predicted by a spin Hamiltonian because it does not include the orbital degree of freedom explicitly. In contrast, the magnetic anisotropy is readily predicted by electronic structure theories employing both orbital and spin degrees of freedom, if one takes into consideration the spin-orbit coupling (SOC), $\lambda \hat{S} \cdot \hat{L}$, of a magnetic ion where $\hat{S}$ and $\hat{L}$ are respectively the spin and orbital operators, and $\lambda$ the SOC constant. It was shown that the preferred spin orientation of a magnetic ion can be predicted and understood in terms of the HOMO-LUMO interactions of the magnetic ion by taking SOC, $\lambda \hat{S} \cdot \hat{L}$, as perturbation. A spin Hamiltonian gives rise to the spin-half misconception, namely, the blind belief that spin-half magnetic ions do not possess magnetic anisotropy that arise from SOC. This misconception contradicts not only experimental observations on spin-half ions but also theoretical results based on DFT calculations and perturbation theory analyses based on an electronic Hamiltonian. This misconception is a direct consequence from the limitedness of a spin Hamiltonian that it lacks the orbital degree of freedom. We show that the magnetic properties of 5d ion oxides are better explained by the LS-coupling than by the jj-coupling scheme of SOC, that the spin-orbital entanglement of 5d ions is not as strong as has been assumed.

# 1. Introduction

In this chapter we examine how to think about and describe the magnetic properties of crystalline solids, which arise from their transition-metal magnetic ions, from the perspectives of an electronic Hamiltonian. The latter represents the energy states of a magnetic system using both orbital and spin degrees of freedom, that is, the angular property of a magnetic ion is described by a set of orbital/spin states $\left| L, L_z \right\rangle \left| S, S_z \right\rangle$. Compared with the strength of chemical bonding (of the order of several eV), the unpaired electrons of a magnetic ion interact very weakly with those of neighboring magnetic ions so that the energy scale involved in magnetic states is very small, and the states responsible for the magnetic properties are closely packed in energy (**Fig. 1**). (For example, at the magnetic field H of 1 Tesla, $\mu_B H = 5.8 \times 10^{-2}$ meV = 0.67 K in $k_B$ units. Other energy scales for discussing magnetic properties are 1 meV = 11.6 K = 8.06 cm$^{-1}$, and 1 cm$^{-1}$ = 1.44 K.) To quantitatively describe the magnetic properties of such a system at any given temperature, it is necessary to obtain the spectrum of the energy states and subsequently Boltzmann-average the properties of these states. Since solving this problem on the basis of an electronic Hamiltonian is very difficult, one employs a spin Hamiltonian $\hat{H}_{spin}$, which represents each magnetic ion using only a set of spin states $\left| S, S_z \right\rangle$. This toy Hamiltonian allows one to generate the energy states without self-consistent-field calculations thereby greatly simplifying calculations, because it is specified by a few spin exchange interactions $J_{ij} \hat{S}_i \cdot \hat{S}_j$ between certain spin sites i and j,

$$\hat{H}_{spin} = \sum_{i<j} J_{ij} \hat{S}_i \cdot \hat{S}_j \qquad (1)$$

where the constants $J_{ij}$ (i.e., spin exchange parameters) are the numerical parameters to be determined. The repeat pattern of the chosen spin exchange paths i-j forms the spin lattice (e.g., an isolated dimer, a uniform chain, an alternating chain, a two-leg ladder, etc.) of the magnetic ions (**Fig. 2**). Once a spin lattice is selected, this model Hamiltonian greatly simplifies the generation of its energy states as a function of the numerical parameters $J_{ij}$, which are fixed as those that best simulate the experimental magnetic data (e.g., magnetic susceptibility, specific heat, and spin wave dispersion relations). The purpose of using such a toy Hamiltonian is to capture the essential physics of observed magnetic properties with a minimal number of adjustable parameters $J_{ij}$.

A general problem facing such a toy Hamiltonian analysis is that more than one spin lattice may equally well simulate the available experimental data. Since the novelty of a chosen spin lattice presents an opportunity to discover a new physics, the practitioners of spin Hamiltonian analyses tend to favor the interpretation of experimental data using a novel spin lattice without checking if the chosen spin lattice is consistent with the electronic structure of a magnetic system under examination. Not infrequently, therefore, a chosen spin lattice turns out to be irrelevant for the system under examination, thus generating "an answer in search of a problem". A bright side of such a regrettable situation would be that the generated physics can stimulate experimental interests to search for a system that fits the "predicted" physics. These days one can readily determine what spin exchanges paths are relevant for any given magnetic system by performing energy-mapping analysis[1-4] on the basis of DFT electronic structure calculations. This theoretical/computational tool makes it possible to interpret experimental data in terms of the relevant spin lattice.

An implicit assumption behind using a spin Hamiltonian is that one can correctly describe all magnetic phenomena in terms of the energy states it generates. The strength of a spin Hamiltonian analysis is to simplify complex calculations as a result of using only the spin degree of freedom, but this strength is also the very cause for its failure at the fundamental level; a spin Hamiltonian description leads to the conceptual impasse recently termed the spin-half syndrome.[5,6] A classic example showing this spin-half misconception is the study of $CuCl_2 \cdot 2H_2O$ by Moriya and Yoshida more than six decades ago;[7] as the cause for the observed spin orientation of the S = 1/2 ion $Cu^{2+}$, they dismissed outright the possibility that the S = 1/2 ion has magnetic anisotropy induced by SOC, $\lambda \hat{S} \cdot \hat{L}$, which is a single-spin site interaction (i.e., a local interaction), and then proceeded to explain the observed spin orientation in terms of nonlocal interactions (e.g., anisotropic spin exchange and magnetic dipole-dipole interactions). Over the years the spin-half misconception has been perpetuated in monographs and textbooks on magnetism.[8] However, this misconception contradicts not only the experimental observations that spin-half ions (e.g., $Cu^{2+}$, $V^{4+}$, $Ir^{4+}$) exhibit magnetic anisotropies,[5,6,9] but also theoretical results based on electronic Hamiltonians in which the energy states of a magnetic system are described by using both orbital and spin degrees of freedom.[5,6,9]

A transition-metal magnetic ion of any spin (S = 1/2 – 5/2) has magnetic anisotropy as a consequence of SOC, $\lambda \hat{S} \cdot \hat{L}$, because the latter induces interactions among its crystal-field split $d$-states and because the energy-lowering associated with these interactions depend on the spin orientation.[2,5,6,9,10] In an electronic Hamiltonian approach the energy states of a magnetic system are discussed in terms of its magnetic orbitals (i.e., its singly occupied orbitals). Each magnetic orbital represents either the up-spin state $\left|\uparrow\right\rangle = \left|\frac{1}{2},+\frac{1}{2}\right\rangle$ or the down-spin state $\left|\downarrow\right\rangle = \left|\frac{1}{2},-\frac{1}{2}\right\rangle$, so the overall spin S of a magnetic ion is related to how many magnetic orbitals it generates. Thus, each magnetic ion of a magnetic orbital in spin state $\left|S,S_z\right\rangle$ ($=\left|\uparrow\right\rangle$ or $\left|\downarrow\right\rangle$ ) is described by the orbital/spin state $\left|L,L_z\right\rangle\left|S,S_z\right\rangle$. The magnetic states are modified by SOC, $\lambda\hat{S}\cdot\hat{L}$, due to the associated intermixing between them, but this intermixing does not occur in the spin part $\left|S,S_z\right\rangle$, but in the orbital part $\left|L,L_z\right\rangle$, of each state. For example, when there is no degeneracy in the magnetic orbitals, a given magnetic orbital $\left|L,L_z\right\rangle\left|S,S_z\right\rangle$ is modified by the intermixing as

$$\left\{(1-\gamma^2-\delta^2-\cdots)\left|L,L_z\right\rangle + \gamma\left|L',L_z'\right\rangle + \delta\left|L'',L_z''\right\rangle + \cdots \right\}\left|S,S_z\right\rangle, \qquad (2)$$

where $\gamma$ and $\delta$ are the mixing coefficients (see Section 7 for more details). This SOC-induced orbital mixing is independent of whether the overall spin S of the magnetic ion is 1/2 or greater because this mixing occurs in each individual magnetic orbital and hence does not depend on how many magnetic orbitals a magnetic ion generates. This is why magnetic anisotropy is predicted for S = 1/2 ions on an equal footing to S > 1/2 ions in an electronic Hamiltonian approach. This fundamental result is not described by a spin Hamiltonian simply because it lacks the orbital degree of freedom; having completely suppressed the orbital $\left|L,L_z\right\rangle$ of a magnetic ion, a spin Hamiltonian does not allow one to discuss the SOC, $\lambda\hat{S}\cdot\hat{L}$, and hence is unable to describe the preferred spin orientation of any magnetic ion. The spin-half misconception is a direct consequence from this deficiency of a spin Hamiltonian.

Anyone who attempted to publish the finding that the spin-half misconception is erroneous would have experienced eye-opening discourses with its proponents (mostly, practitioners of spin Hamiltonian analyses), to learn that they treat the attempt as an affront to their work and do their utmost to suppress its publication. For those schooled in the electronic structure description, it is only natural to describe the energy states of a magnetic system by using both orbital and spin degrees of freedom, because unpaired electrons responsible for magnetic properties must be accommodated in certain orbitals, and hence have no problem in finding that a spin Hamiltonian is a theoretically limited tool. However, most of those schooled in doing physics with spin Hamiltonian do not appear to realize that this toy Hamiltonian was born out of the necessity to simplify calculations. They tend to believe that the correct energy states of a magnetic system are those generated by using only the spin degree of freedom, and insist that an electronic Hamiltonian description should produce the same conclusion as does a spin Hamiltonian description even if it is an erroneous one resulting from its deficiency. To help break this conceptual impasse, it is necessary to expose the origin of the spin-half misconception by discussing how the properties of solid state magnetic materials are described from the perspectives of an electronic Hamiltonian.

Analysis of magnetic properties on the basis of an electronic Hamiltonian deals with two competing issues; one is to produce accurate quantitative predictions, and the other is to provide qualitative pictures with which to organize and think about. These two subjects are discussed by organizing our work as follows: In Section 2 we first discuss the angular properties of the atomic orbitals and then the crystal-field split $d$-states of magnetic ions. Section 3 covers the energy-mapping analysis that allows one to relate the spin Hamiltonian analysis of a given magnetic system to its electronic structure by evaluating the spin exchange parameters this toy Hamiltonian needs. In Section 4 we discuss the qualitative features of spin exchange interactions in terms of orbital interactions. In Section 5 we describe indirect ways of incorporating SOC into a spin Hamiltonian and the associated energy-mapping analysis as well as the origin of the spin-half misconception. The condition leading to uniaxial magnetism is discussed in Section 6 to prepare for our discussion of magnetic anisotropy. Section 7 describes the qualitative rules that allow one to predict the preferred spin orientations of magnetic ions on the basis of the perturbation theory in which SOC is taken as perturbation with the crystal-field split *d*-states as the unperturbed state. In Section 8 we discuss several issues concerning the magnetic properties of 5d magnetic ions. Our concluding remarks are summarized in Section 9.

## 2. Atomic orbitals and magnetic orbitals

### 2.1. Angular properties of atomic orbitals

The angular properties of atomic orbitals are specified by the spherical harmonics, $\left| L, L_z \right\rangle$, defined in terms of two quantum numbers; the orbital quantum number L (= 0, 1, 2, … ) and its z-axis component $L_z = -L, -L+1, …, L-1, L$ for a given L. The angular behaviors of the atomic *p*- and *d*-orbitals are summarized in **Table 1**. In terms of the magnetic quantum numbers $L_z$, the *d*-orbitals are grouped into three sets:

$$L_z = 0 \quad \text{for } 3z^2 - r^2$$
$$L_z = \pm 1 \quad \text{for } \{xz, yz\}$$
$$L_z = \pm 2 \quad \text{for } \{xy, x^2 - y^2\}$$

Similarly, the p-orbitals are expressed as linear combinations of the spherical harmonics $\left| L, L_z \right\rangle$, where L =1, and $L_z$ = 1, 0, −1. Thus, the *p*-orbitals are grouped into two sets:

$$L_z = 0 \quad \text{for} \quad z$$
$$L_z = \pm 1 \quad \text{for} \quad \{x, y\}$$

Consequently, as depicted in **Fig. 3**, the minimum difference $|\Delta L_z|$ in the magnetic quantum numbers between different atomic orbitals is summarized as follows:

$$|\Delta L_z| = 0 \;\; \text{between} \begin{cases} xz \text{ and } yz \\ xy \text{ and } x^2 - y^2 \\ x \text{ and } y \end{cases}$$

$$|\Delta L_z| = 1 \;\; \text{between} \begin{cases} 3z^2 - r^2 \text{ and } \{xz, yz\} \\ \{xz, yz\} \text{ and } \{xy, x^2 - y^2\} \\ z \text{ and } \{x, y\} \end{cases}$$

$$|\Delta L_z| = 2 \;\; \text{between } 3z^2\text{-}r^2 \text{ and } \{xy, x^2\text{-}y^2\}$$

These $|\Delta L_z|$ values play a crucial role in understanding the preferred spin orientations of magnetic ions on the basis of the SOC-induced HOMO-LUMO interactions of their crystal-field split *d*-states (see Section 7).

In quantum mechanics the orbital angular momentum $\vec{L}$ is replaced by the orbital angular momentum operator $\hat{L}$, which has three components $\hat{L}_x, \hat{L}_y$ and $\hat{L}_z$ in a Cartesian coordinate system. Most calculations associated with orbital angular momentum make use of $\hat{L}_z$, $\hat{L}_+$ and $\hat{L}_-$, where $\hat{L}_+$ and $\hat{L}_-$ are the ladder operators defined by

$$\hat{L}_+ = \hat{L}_x + i\hat{L}_y$$
$$\hat{L}_- = \hat{L}_x - i\hat{L}_y$$

The orbitals $|L, L_z\rangle$ are affected by the operators $\hat{L}_z$, $\hat{L}_+$ and $\hat{L}_-$ as follows:

$$\hat{L}_z \big| L, L_z \big\rangle = L_z \big| L, L_z \big\rangle$$
$$\hat{L}_+ \big| L, L_z \big\rangle = \sqrt{L(L+1) - L_z(L_z+1)} \, \big| L, L_z + 1 \big\rangle \qquad (3)$$
$$\hat{L}_- \big| L, L_z \big\rangle = \sqrt{L(L+1) - L_z(L_z-1)} \, \big| L, L_z - 1 \big\rangle$$

Here we use the atomic unit in which the unit of angular momentum, $\hbar$, is equal to 1. The $\hat{L}_+$ raises the $L_z$ of $\big| L, L_z \big\rangle$ by 1 as long as $L_z + 1 \leq L$, while $\hat{L}_-$ lowers the $L_z$ of $\big| L, L_z \big\rangle$ by 1 as long as $L_z - 1 \geq -L$. In our later discussion, we need to evaluate the integrals

$$\big\langle i \big| \hat{L}_x \big| j \big\rangle, \ \big\langle i \big| \hat{L}_y \big| j \big\rangle \ \text{and} \ \big\langle i \big| \hat{L}_z \big| j \big\rangle$$

involving atomic $p$-orbitals $(i, j = x, y, z)$ as well as those involving atomic $d$-orbitals $(i, j = 3z^2 - r^2, xz, yz, x^2 - y^2, xy)$. By using Eq. 3 and the expressions of the atomic orbitals listed in **Table 1**, we obtain the nonzero integrals listed in **Table 2**.[9]

## 2.2. Crystal-field split d-states

In most cases we are concerned with systems containing transition-metal magnetic ions M in magnetic oxides. The preferred orientations of their spin moments are determined by their $d$-states split by their surrounding ligands L. It depends on the symmetry and composition of the $ML_n$ (typically, n = 4 − 6) polyhedron how the $d$-states of the ion M split. In a description of electronic structures using an effective one-electron Hamiltonian $\hat{H}^{\text{eff}}$, each split $d$-level of a $ML_n$ polyhedron does not change its energy and shape regardless of whether it is occupied by one or two electrons, because the presence of electron-electron repulsion in a doubly-occupied level is ignored. We discuss this simple picture first and then consider how to modify these one-electron levels by electron correlation.

### 2.2.1. One-electron states without electron correlation

How strongly the $d$-orbitals of the transition metal $M$ interact with the $p$-orbitals of the ligands L depends on the nature of the $d$-orbitals and the shape of the $ML_n$ polyhedron.[11] In the split $d$-states that result from these interactions, the ligand $p$-orbitals are combined out-of-phase to the metal $d$-orbitals. Therefore, a given $d$-state lies high in energy if the $M$-$L$ antibonding is strong in the state. Let us start from the $d$-states of an $ML_6$ octahedron (**Fig. 4a**), which are split into the triply-degenerate $t_{2g}$ state lying below the doubly-degenerate $e_g$ state (**Fig. 4b**). The three components of the $t_{2g}$ state are each described by $M$-$L$ $\pi$-antibonding, and the two components of the $e_g$ state by $M$-$L$ $\sigma$-antibonding (**Fig. 4c**). Some $ML_n$ (typically, n = 4 – 6) polyhedra can be regarded as derived from the $ML_6$ octahedron by lengthening and/or removing a few M-L bonds. The split $d$-states of such polyhedra can be readily predicted by considering how the extent of the $\sigma$-antibonding and/or $\pi$-antibonding of the $M$-$L$ bonds varies under the geometrical changes (**Fig. 5**).

For an axially-elongated $ML_6$ octahedron with the z-axis taken along the elongated $M$-$L$ bonds, the $d$-states are split as depicted in **Fig. 5b**; the $3z^2$–$r^2$ state (commonly, referred to as the $z^2$ state, for simplicity) is significantly lowered in energy because the $\sigma$-antibonding is reduced, while the xz and yz states are slightly lowered in energy because the $\pi$-antibonding is reduced. For a square-planar $ML_4$ with the z-axis taken perpendicular to the plane, the $d$-states are split as shown in **Fig. 5c**; the $3z^2$–$r^2$ state is lowered to become the lowest in energy because the $\sigma$-antibonding along the z-direction is totally absent while that in the xy-plane is further reduced because the girdle of the $3z^2-r^2$ state is diminished in size by the second-order orbital mixing of the upper $s$-orbital of $M$.[11] In addition, the xz and yz states of the $ML_4$ square plane are lower than those of the axially-elongated $ML_6$ octahedron because the $\pi$-antibonding is absent along the z-direction. For a linear $ML_2$ with the z-axis taken along the $M$-$L$ bonds, the $d$-states are split as depicted in **Fig. 5d**; the xy and $x^2-y^2$ states are lowered more in energy than are the xz and yz states because $\pi$-antibonding is absent in the xy and $x^2-y^2$ states while it is present in the xz and yz states.

In discussing the $t_{2g}$ and $e_g$ states of an $ML_6$ octahedron, there occur cases when it is more convenient to take the z-axis along one 3-fold rotational axis of the octahedron (**Fig. 6a**) [12] rather than along one $M$-$L$ bond (i.e., along one 4-fold rotational axis) (**Fig. 5a**). Then their orbital character changes as summarized in **Table 3**; the $3z^2-r^2$ state becomes one of the $t_{2g}$ set, while the (xy, $x^2-y^2$) degenerate set mixes with the (xz, yz) degenerate set to give the ($1e_x$, $1e_y$) and ($2e_x$, $2e_y$) sets (**Fig. 6b**). The (xy, $x^2-y^2$) set has a larger contribution than does the (xz, yz) set in the ($1e_x$, $1e_y$) set, and the opposite is the case in the ($2e_x$, $2e_y$) set (**Table 3**). Such orbital representations as described by **Fig. 6** and **Table 3** will be employed in Section 7.

### 2.2.2. One-electron states with electron correlation

The essence of electron correlation is that, when a given energy state is doubly occupied, its energy is raised by electron-electron repulsion. The latter is partly reduced in spin-polarized electronic structure calculations, in which up-spin states are allowed to differ in energy and shape from their down-spin counterparts. For strongly correlated systems, the energy split arising from spin-polarized electronic structure calculations is not strong enough to generate singly-occupied states needed to describe their magnetic insulating states. In spin-polarized DFT calculations, this deficiency is corrected by adding the on-site repulsion U on magnetic ions to force a large split between their up-spin and down-spin states (**Fig. 7**).[13] Such calculations are referred to as DFT+U calculations.

An important consequence of spin polarized DFT+U calculations is found when two adjacent spin sites interact.[2] If the two equivalent spin sites have a ferromagnetic (FM) arrangement (**Fig. 8a**), the up-spin states of the two sites are degenerate, and so are the down-spin states of the two sites. However, if the two equivalent spin sites have an antiferromagnetic (AFM) arrangement (**Fig. 8b**), the up-spin states of the two sites are nondegenerate, and so are the down-spin states of the two sites. In general, orbital interactions between degenerate states are stronger than those between nondegenerate states.[11] Since orbital interactions between states require that their spins be identical, the AFM arrangement leads to a weaker orbital interaction between adjacent spin sites than does the FM arrangement.[2]

From the viewpoint of the split $d$-states obtained from an effective one-electron Hamiltonian, the qualitative features of DFT+U calculations can be simulated by splitting the up-spin $d$-states from those of the down-spin $d$-states approximately by the amount of U, as illustrated in **Fig. 9**, for a high-spin (S = 2) $d^6$ ion forming a square planar site forming a $FeL_4$ square plane. For simplicity, the separation between the up-spin and down-spin $d$-states is exaggerated in **Fig. 9**. What is important to note is that the HOMO and the LUMO levels occur within the down-spin states if the $d$-shell is more than half-filled, but within the up-spin states if the $d$-shell less than half-filled. (This is due to the convention in which the majority and minority spin states are regarded as up-spin and down-spin states, respectively.) Only when the $d$-shell is half-filled in a high-spin manner, the HOMO and the LUMO levels occur between the up-spin and down-spin states.

An alternative way of correcting the deficiency of spin-polarized DFT calculations is the hybrid functional method,[14] in which the exchange-correlation functional needed for calculations is obtained by mixing some amount, $\alpha$ (typically, 0.2), of the Hartree-Fock exchange potential into the DFT functional. The on-site repulsion U is an empirical parameter in DFT+U calculations, and so is the mixing parameter $\alpha$ in DFT+hybrid calculations. In general, DFT+U calculations are much less time-consuming than are DFT+hybrid calculations. It should be emphasized that density functional calculations are first principles calculations only after the value of U is fixed in DFT+U calculations, and only after the value of $\alpha$ is fixed in DFT+hybrid calculations.

Given computing resources, DFT calculations with or without including SOC effects[15] can be readily carried out by using user-friendly DFT program packages such as VASP,[16] which considers only valence electrons using the frozen-core projector augmented waves, and WIEN2k,[17] which considers all electrons. As the exchange-correlation functional needed for DFT calculations, the generalized gradient approximation [18] is commonly used for studying solid state materials. In understanding results of DFT, DFT+U and DFT+U+SOC calculations or predicting results prior to calculations, the concept of orbital interaction analysis,[11] developed on the basis of one electron picture, is useful (see below).

### 3. Energy mapping analysis

For two spins $\hat{S}_1$ and $\hat{S}_2$ at spin sites 1 and 2, respectively, the dot product $\hat{S}_1 \cdot \hat{S}_2$ has three Cartesian components, i.e., $\hat{S}_1 \cdot \hat{S}_2 = \hat{S}_{1x}\hat{S}_{2x} + \hat{S}_{1y}\hat{S}_{2y} + \hat{S}_{1z}\hat{S}_{2z}$. Thus a general expression for the spin exchange interaction energy between the two spin sites can be written as

$$\hat{H}_{spin} = J_x\hat{S}_{1x}\hat{S}_{2x} + J_y\hat{S}_{1y}\hat{S}_{2y} + J_z\hat{S}_{1z}\hat{S}_{2z},$$

where $J_x$, $J_y$ and $J_z$ are anisotropic spin exchanges along the x-, y- and z-directions, respectively. If $J_x = J_y = J_z = J$, namely, if the spin exchange is isotropic, the above expression is simplified as

$$\hat{H}_{spin} = J(\hat{S}_{1x}\hat{S}_{2x} + \hat{S}_{1y}\hat{S}_{2y} + \hat{S}_{1z}\hat{S}_{2z}),$$

which represents a Heisenberg spin Hamiltonian. Another extreme case is given by $J_x = J_y = 0$, for which we obtain an Ising spin Hamiltonian

$$\hat{H}_{spin} = J_z\hat{S}_{1z}\hat{S}_{2z}.$$

This Hamiltonian describes a magnetic system made up of uniaxial magnetic ions (namely, those ions with a nonzero moment only in one direction, see Section 6). The deviation of spin exchange from the isotropic character is a consequence of SOC. In this section we focus on how to determine isotropic spin exchanges, which are often referred to as Heisenberg or symmetric spin exchanges. The evaluation of anisotropic spin exchanges will be discussed in Section 5.2.

### 3.1. Use of eigenstates for an isolated spin dimer[1,19]

To gain insight into the meaning of the spin exchange interaction, we consider a spin dimer consisting of two equivalent spin-1/2 spin sites, 1 and 2, with one electron at each spin site (**Fig. 10**). The energy of the spin dimer arising from the spin exchange interaction between the spins $\hat{S}_1$ and $\hat{S}_2$ is given by the spin Hamiltonian

$$\hat{H}_{spin} = J\,\hat{S}_1 \cdot \hat{S}_2, \tag{4a}$$

where J is the spin exchange parameter. If the spins are regarded as vectors $\vec{S}_1$ and $\vec{S}_2$, then the Hamiltonian is written as

$$\hat{H}_{spin} = J\,\vec{S}_1 \cdot \vec{S}_2 \tag{4b}$$

In the present work, we will use the operator and vector representations of spin interchangeably. Note the absence of the negative sign in this expression. With this definition, the AFM and FM spin exchange interactions are given by J > 0 and J < 0, respectively. Given the dot product between $\vec{S}_1$ and $\vec{S}_2$, the lowest energy for J > 0 occurs when the angle θ between the two spins is 180° (i.e., the spins are AFM), but that for J < 0 when θ = 0° (i.e., the spins are FM). In either case, the spin Hamiltonian leads to a collinear spin arrangement.

In principle, the spin at site i (= 1, 2) of the spin dimer can have either up-spin $\left|\uparrow\right\rangle$ or down-spin $\left|\downarrow\right\rangle$ state. For a single spin S = 1/2 and $S_z = \pm 1/2$ so that, in terms of the $\left|S.S_z\right\rangle$ notations, these states are given by

$$\begin{aligned}\left|\uparrow\right\rangle &= \left|\tfrac{1}{2}, +\tfrac{1}{2}\right\rangle \\ \left|\downarrow\right\rangle &= \left|\tfrac{1}{2}, -\tfrac{1}{2}\right\rangle\end{aligned}.$$

These states obey the following general relationships:

$$\hat{S}_z \left| S, S_z \right\rangle = S_z \left| S, S_z \right\rangle$$
$$\hat{S}_+ \left| S, S_z \right\rangle = \sqrt{S(S+1) - S_z(S_z+1)} \left| S, S_z + 1 \right\rangle \qquad (5)$$
$$\hat{S}_- \left| S, S_z \right\rangle = \sqrt{S(S+1) - S_z(S_z-1)} \left| S, S_z - 1 \right\rangle$$

where the ladder operators are given by

$$\hat{S}_+ = \hat{S}_x + i\,\hat{S}_y$$
$$\hat{S}_- = \hat{S}_x - i\,\hat{S}_y$$

Using these ladder operators, Eq. 4a is rewritten as

$$\hat{H}_{spin} = J\,\hat{S}_{1z}\hat{S}_{2z} - J\,(\hat{S}_{1+}\hat{S}_{2-} + \hat{S}_{1-}\hat{S}_{2+})\,/\,2 \qquad (4c)$$

The eigenstates of $\hat{H}_{spin}$ allowed for the spin dimer are the singlet state $\left| S \right\rangle$ and triplet state $\left| T \right\rangle$, which are given by

$$\left| S \right\rangle = \left( \left| \uparrow \right\rangle_1 \left| \downarrow \right\rangle_2 - \left| \downarrow \right\rangle_1 \left| \uparrow \right\rangle_2 \right) / \sqrt{2}$$

$$\left| T \right\rangle = \begin{cases} \left| \uparrow \right\rangle_1 \left| \uparrow \right\rangle_2 \\ \left| \downarrow \right\rangle_1 \left| \downarrow \right\rangle_2 \\ \left( \left| \uparrow \right\rangle_1 \left| \downarrow \right\rangle_2 + \left| \downarrow \right\rangle_1 \left| \uparrow \right\rangle_2 \right) / \sqrt{2} \end{cases}$$

Note that the broken-symmetry (or Néel) states,

$$\left| \uparrow \right\rangle_1 \left| \downarrow \right\rangle_2 \text{ and } \left| \downarrow \right\rangle_1 \left| \uparrow \right\rangle_2,$$

interact through $\hat{H}_{spin}$ to give the symmetry-adapted states $\left| S \right\rangle$ and $\left| T \right\rangle$. We evaluate the energies of $\left| T \right\rangle$ and $\left| S \right\rangle$ by using Eq. 5 to find

$$E_{spin}(T) = \left\langle T \left| \hat{H}_{spin} \right| T \right\rangle = J/4$$

$$E_{spin}(S) = \left\langle S \left| \hat{H}_{spin} \right| S \right\rangle = -3J/4.$$

Thus, the energy difference between the two states is given by

$$\Delta E_{spin} = E_{spin}(T) - E_{spin}(S) = J, \tag{6}$$

so the spin exchange constant J represents the energy difference between the singlet and triplet spin states of the spin dimer. The singlet state is lower in energy than the triplet state if the spin exchange J is AFM (i.e., $J > 0$), and the opposite is the case if the spin exchange J is FM (i.e., $J < 0$).

We now examine how the triplet and singlet states of the spin dimer are described in terms of electronic structure calculations. The electronic Hamiltonian $\hat{H}_{elec}$ for this two-electron system can be written as

$$\hat{H}_{elec} = \hat{h}(1) + \hat{h}(2) + 1/r_{12}, \tag{7}$$

where $\hat{h}(i)$ ($i = 1, 2$) is the one-electron energy (i.e., the kinetic and the electron-nuclear attraction energies) of the electron i ($= 1, 2$), and $r_{12}$ is the distance between electrons 1 and 2. Assume that the unpaired electrons at sites 1 and 2 are accommodated in the orbitals $\phi_1$ and $\phi_2$, respectively, in the absence of interaction between them. Such singly-occupied orbitals are referred to as magnetic orbitals. The weak interaction between $\phi_1$ and $\phi_2$ leads to the two levels $\psi_1$ and $\psi_2$ of the dimer separated by a small energy gap $\Delta e$ (**Fig. 11**), which are approximated by

$$\psi_1 = (\phi_1 + \phi_2)/\sqrt{2}$$
$$\psi_2 = (\phi_1 - \phi_2)/\sqrt{2}.$$

As depicted in **Fig. 12**, one of the three triplet-state wave functions is represented by the electron configuration $\Psi_T$. When $\Delta e$ is very small (compared with that expected for chemical bonding), the singlet state electron configurations $\Phi_1$ and $\Phi_2$ are very close in energy, and interact strongly under $\hat{H}_{elec}$ to give

$$\langle \Phi_1 | \hat{H}_{elec} | \Phi_2 \rangle = K_{12},$$

where $K_{12}$ is the exchange repulsion between $\phi_1$ and $\phi_2$.

$$K_{12} = \langle \phi_1(1)\phi_2(2) | 1/r_{12} | \phi_2(1)\phi_1(2) \rangle,$$

which is the self-repulsion resulting from the overlap density $\phi_1\phi_2$. Thus the true singlet state $\Psi_S$ is described by the lower-energy state of the configuration-interaction (CI) wave functions $\Psi_i$ (i = 1, 2),

$$\Psi_i = C_{1i}\Phi_1 + C_{2i}\Phi_2 \quad (i = 1, 2),$$

namely, $\Psi_S = \Psi_1$. The energies of $\Psi_S$ and $\Psi_T$, $E_{CI}(S)$ and $E_{CI}(T)$, respectively, can be evaluated in terms of $\hat{H}_{elec}$ by using the dimer orbitals $\psi_1$ and $\psi_2$ determined from the calculations for the triplet state $\Psi_T$. Then, after some manipulations, the electronic energy difference between the singlet and triplet state is written as[1,19]

$$\Delta E_{CI} = E_{CI}(S) - E_{CI}(T) = -2K_{12} + \frac{(\Delta e)^2}{U}. \qquad (8)$$

The effective on-site repulsion U is given by

$$U = J_{11} - J_{12},$$

where $J_{11}$ and $J_{12}$ are the Coulomb repulsions

$$J_{11} = \langle \phi_1(1)\phi_1(2) | 1/r_{12} | \phi_1(1)\phi_1(2) \rangle$$
$$J_{12} = \langle \phi_1(1)\phi_2(2) | 1/r_{12} | \phi_1(1)\phi_2(2) \rangle.$$

Then, by mapping the energy spectrum of $\hat{H}_{spin}$ onto that of $\hat{H}_{elec}$, namely,

$\Delta E_{spin} = \Delta E_{CI}$,

we obtain

$$J = \Delta E_{CI} = -2K_{12} + \frac{(\Delta e)^2}{U} \tag{9}$$

It is important to note the qualitative aspect of the spin exchange J on the basis of the above expression. Since the repulsion terms $K_{12}$ and U are always positive, the spin exchange J is divided into the FM and AFM components $J_F$ ($< 0$) and $J_{AF}$ ($> 0$), respectively. That is,

$J = J_F + J_{AF}$,

where

$$J_F = -2K_{12} \tag{10a}$$

$$J_{AF} = \frac{(\Delta e)^2}{U} \tag{10b}$$

The FM term $J_F$ term becomes stronger with increasing the exchange integral $K_{12}$, which in turn increases with increasing the overlap density, $\phi_1\phi_2$. The AFM term $J_{AF}$ becomes stronger with increasing $\Delta e$, which in turn becomes larger with increasing the overlap integral, $\langle \phi_1 | \phi_2 \rangle$. In addition, the $J_{AF}$ term becomes weaker with increasing the on-site repulsion, U.

## 3.2. Use of broken-symmetry states for an isolated spin dimer

For a general magnetic system, it is practically impossible to determine the eigenvalue spectrum of either $\hat{H}_{elec}$ or $\hat{H}_{spin}$. However, for broken-symmetry states, which are not eigenstates of $\hat{H}_{elec}$ and $\hat{H}_{spin}$, their relative energies can be readily determined in terms of both $\hat{H}_{elec}$ and $\hat{H}_{spin}$. With DFT calculations, the energy-mapping for a spin dimer between the energy spectra of $\hat{H}_{elec}$ and $\hat{H}_{spin}$ is carried out by using high-spin and broken-symmetry states ($\left|HS\right\rangle$ and $\left|BS\right\rangle$, respectively).[1-5,20,21] For example, let us reconsider the spin dimer shown in **Fig. 10**, for which the pure-spin $\left|HS\right\rangle$ and $\left|BS\right\rangle$ states are given by

$$\left|HS\right\rangle = \left|\uparrow\right\rangle_1 \left|\uparrow\right\rangle_2 \; \text{or} \; \left|\downarrow\right\rangle_1 \left|\downarrow\right\rangle_2$$
$$\left|BS\right\rangle = \left|\uparrow\right\rangle_1 \left|\downarrow\right\rangle_2 \; \text{or} \; \left|\downarrow\right\rangle_1 \left|\uparrow\right\rangle_2$$

Here the $\left|HS\right\rangle$ state is an eigenstate of the spin Hamiltonian $\hat{H}_{spin}$ in Eq. 3a, but the $\left|BS\right\rangle$ state is not. In terms of this Hamiltonian, the energies of the collinear-spin states $\left|HS\right\rangle$ and $\left|BS\right\rangle$ are given by

$$E_{spin}(HS) = \left\langle HS\right|\hat{H}_{spin}\left|HS\right\rangle = J/4$$
$$E_{spin}(BS) = \left\langle BS\right|\hat{H}_{spin}\left|BS\right\rangle = -J/4,$$

Thus,

$$\Delta E_{spin} = E_{spin}(HS) - E_{spin}(BS) = J/2.$$

In terms of DFT calculations, the electronic structures of the $\left|HS\right\rangle$ and $\left|BS\right\rangle$ states are readily evaluated to determine their energies, $E_{DFT}(HS)$ and $E_{DFT}(HS)$, respectively, and hence obtain the energy difference

$$\Delta E_{DFT} = E_{DFT}(HS) - E_{DFT}(BS).$$

Consequently, by mapping $\Delta E_{spin}$ onto $\Delta E_{DFT}$, we obtain

$$J/2 = \Delta E_{DFT}. \tag{11}$$

### 3.3. Use of broken-symmetry states for general magnetic solids

The energy-mapping analysis based on DFT calculations employs the broken-symmetry state that is not an eigenstate of the spin Hamiltonian. For a general spin Hamiltonian defined in terms of several spin exchange parameters (Eq. 1), it is impossible to determine its eigenstates analytically in terms of the spin exchange parameters to be determined and is also difficult to determine them numerically even when their values are known. For any realistic magnetic system requiring a spin Hamiltonian defined in terms of various spin exchange parameters, the energy-mapping analysis based on DFT greatly facilitates the quantitative evaluation of the spin exchange parameters because it does not rely on the eigenstates but on the broken-symmetry states of the spin Hamiltonian. For broken-symmetry states, the energy expressions of the spin Hamiltonian can be readily written down (see below) and the corresponding electronic energies can be readily determined by DFT calculations as well.

In general, the magnetic energy levels of a magnetic system are described by employing a spin Hamiltonian $\hat{H}_{spin}$ defined in terms of several different spin exchange parameters (Eq. 1). This model Hamiltonian generates a set of magnetic energy levels as the sum of pair-wise interactions $J_{ij} \hat{S}_i \cdot \hat{S}_j$. It is interesting that the sum of such "two-body interactions" can reasonably well describe the magnetic energy spectrum. This is due to the fact that spin exchange interactions are determined primarily by the tails of magnetic orbitals (see Section 4).[1,2] The spin exchange constants $J_{ij}$ of a given magnetic system can be evaluated by employing the energy-mapping method as described below.[2]

(a) Select a set of $N$ spin exchange paths $J_{ij}$ (= $J_1$, $J_2$, … , $J_N$) for a given magnetic system on the basis of inspecting the geometrical arrangement of its magnetic ions and also considering the nature of its M-L-M and M-L…L-M exchange paths.

(b) Construct $N+1$ ordered spin states (i.e., broken-symmetry states) i = 1, 2, … , $N+1$, in which all spins are collinear so that any given pair of spins has either FM or AFM arrangement. For a general spin dimer whose spin sites i and j possess $N_i$ and $N_j$ unpaired spins (hence, spins $S_i = N_i/2$ and $S_j = N_j/2$), respectively, the spin exchange energies of the FM and AFM arrangements ($E_{FM}$ and $E_{AFM}$, respectively) are given by[3]

$$E_{FM} = +N_i N_j J_{ij}/4 = +S_i S_j J_{ij},$$
$$E_{AFM} = -N_i N_j J_{ij}/4 = -S_i S_j J_{ij}, \qquad (12)$$

where $J_{ij}$ (= $J_1$, $J_2$, … , $J_N$) is the spin exchange parameter for the spin exchange path ij = 1, 2, … , $N$. Thus, the total spin exchange energy of an ordered spin arrangement is readily obtained by summing up all pair-wise interactions to find the energy expression $E_{spin}(i)$ (i = 1, 2, … , $N+1$) in terms of the parameters to be determined and hence the $N$ relative energies

$$\Delta E_{spin}(i-1) = E_{spin}(i) - E_{spin}(1) \qquad (i = 2, 3, … , N+1)$$

(c) Determine the electronic energies $E_{DFT}(i)$ of $N+1$ ordered spin states i = 1, 2, … , $N+1$ by DFT calculations to obtain the $N$ relative energies

$$\Delta E_{DFT} (i - 1) = E_{DFT}(i) - E_{DFT}(1) \qquad (j = 2, 3, \ldots, N+1)$$

As already mentioned, DFT calculations for a magnetic insulator tend to give a metallic electronic structure because the electron correlation of a magnetic ion leading to spin polarization is not well described. Since we deal with the energy spectrum of a magnetic insulator, it is necessary that the electronic structure of each ordered spin state obtained from DFT calculations be magnetic insulating. To ensure this aspect, it is necessary to perform DFT+U calculations [13] by adding on-site repulsion $U_{eff} = U - J$ with on-site repulsion U and on-site exchange interaction J on magnetic ions. Furthermore, as can be seen from Eq. 10b, the AFM component of a spin exchange decreases with increasing $U_{eff}$ so that the magnitude and sign of a spin exchange constant may be affected by $U_{eff}$. It is therefore necessary to carry out DFT+U calculations with several different $U_{eff}$ values.

(d) Finally, determine the values of $J_1$, $J_2$, $\ldots$, $J_N$ by mapping the *N* relative energies $\Delta E_{DFT}$ onto the *N* relative energies $\Delta E_{spin}$,

$$\Delta E_{DFT} (i - 1) = \Delta E_{spin}(i - 1) \qquad (i = 2 - N+1) \qquad (13)$$

In determining *N* spin exchanges $J_1$, $J_2$, $\ldots$, $J_N$, one may employ more than *N*+1 ordered spin states, hence obtaining more than *N* relative energies $\Delta E_{DFT}$ and $\Delta E_{spin}$ for the mapping. In this case, the *N* parameters $J_1$, $J_2$, $\ldots$, $J_N$ can be determined by performing least-squares fitting analysis.

### 3.4. Energy-mapping based on four ordered spin states[4]

For our calculations, we regard the spin operators $\hat{S}_i$ and $\hat{S}_j$ as the classical vectors of $\vec{S}_i$ and $\vec{S}_j$, respectively. Then, the spin Hamiltonian can be written as

$$\hat{H}_{spin} = \sum_{i<j} J_{ij} \hat{S}_i \cdot \hat{S}_j \rightarrow \sum_{i<j} J_{ij} \vec{S}_i \cdot \vec{S}_j \qquad (14)$$

Without loss of generality, the spin pair i-j will be regarded as 1-2. For simplicity, all spin sites are assumed to have an identical spin S. We carry out DFT+U calculations for the following four ordered spin states:

| State | Spin 1 | Spin 2 | Other spin sites |
|-------|--------|--------|------------------|
| 1 | (0, 0, S) | (0, 0, S) | Either (0, 0, S) or (0, 0, -S) |
| 2 | (0, 0, S) | (0, 0, -S) | according to the experimental (or a |
| 3 | (0, 0, -S) | (0, 0, S) | low-energy) spin state. Keep the |
| 4 | (0, 0, -S) | (0, 0, -S) | same for the four spin states |

where the notations (0, 0, S) and (0, 0, -S), for example, mean that the spin vectors are pointed along the positive and negative z-directions, respectively. We represent the energies of the spin states 1 – 4 as $E_1$ – $E_4$, respectively. Then, according to Eq. 14, the energy difference, $E_1 + E_4 - E_2 - E_3$, is related to the spin exchange J as

$$J_{12} = \frac{E_1 + E_4 - E_2 - E_3}{4S^2} \qquad (15)$$

Once the energies $E_1$ – $E_4$ are obtained from DFT+U+SOC calculations, we can readily determine $J_{12}$.

## 3.5. General features of spin exchanges numerically extracted

Common DFT functionals suffer from the self-interaction error, i.e., a single electron interacts with itself, which is unphysical. This error results in a spurious delocalization of orbitals including magnetic orbitals. Consequently, spin exchange interactions are overestimated by the usual DFT methods. This self-interaction error can be reduced by using the DFT+U method, in which the on-site Coulomb interaction is taken into consideration. This on-site interaction is parameterized by the effective on-site Coulomb interaction $U_{eff} = U - J$. By adding such Hartree-Fock-like terms, the DFT+U method makes the magnetic orbitals more localized and decreases the overlap between magnetic orbitals hence reducing the magnitudes of spin exchange interactions. Currently, there is no reliable way of determining the U and J parameters needed for DFT+U calculations. A practical way of probing the magnetic properties of a given system is to carry out DFT+U calculations for several different $U_{eff}$ values, which provide several sets of the $J_1$, $J_2$, … , $J_N$ values. It is important to find trends common to these sets. What matters in finding a spin lattice are the relative magnitudes of the spin exchanges. As already pointed out, the purpose of using a spin Hamiltonian is to quantitatively describe the observed experimental data with a minimal set of $J_{ij}$ values hence capturing the essence of the chemistry and physics involved. Experimentally, such a set of $J_{ij}$ values for a given magnetic system is deduced first by choosing a few spin exchange paths $J_{ij}$ that one considers as important for the system and then by evaluating their signs and magnitudes such that the energy spectrum of the resulting spin Hamiltonian best simulates the observed experimental data. The numerical values of $J_{ij}$ deduced from this fitting analysis depends on what spin lattice model one employs for the fitting, and hence more than one spin lattice may fit the experimental data equally well. This non-uniqueness of the fitting analysis has been the source of controversies in the literature over the years. Ultimately, the spin lattice of a magnetic system deduced from experimental fitting analysis should be consistent with the one determined from the energy-mapping analysis based on DFT calculations, because the observed magnetic properties are a consequence of the electronic structure of the magnetic system.

## 4. Orbital interactions controlling spin exchanges

For a given magnetic system, one can determine the values of its various spin exchanges using the energy mapping analysis based on DFT+U calculations and hence ultimately find the spin lattice appropriate for it. What the energy-mapping analysis cannot tell us is why a certain spin exchange is strong or weak. To answer this question, it is necessary to understand how the strength of a given spin exchange interaction between two magnetic ions is related to the orbital interaction between the magnetic orbitals representing the magnetic ions. In this section, we consider the qualitative aspects of the orbital interactions controlling spin exchange interactions.

Given a magnetic solid made up of $ML_n$ polyhedra containing a magnetic transition cation $M^{x+}$ (x = oxidation state), there may occur two types of spin exchange paths, namely, $M$-$L$-$M$ exchange and/or $M$-$L$...$L$-$M$ exchange paths. The qualitative factors governing the signs and magnitudes of $M$-$L$-$M$ exchanges were well established many decades ago.[22,23] However, the importance of $M$-$L$...$L$-$M$ exchange paths has been realized much later.[1,2] In leading to AFM interactions, $M$-$L$...$L$-$M$ exchanges can be much stronger than $M$-$L$-$M$ exchanges. What was not realized in the early studies of M-L-M exchanges is the importance of the magnetic orbitals of $ML_n$ polyhedra, in which the $M$ d-orbitals are combined out-of-phase with the $L$ p-orbitals. In $M$-$L$...$L$-$M$ spin exchanges the magnetic orbitals of the two metal sites can interact strongly as long as their $L$ p-orbital tails can interact through the *L...L* contact.[1] In what follows we examine qualitatively the through-space and through-bond orbital interactions[2] that govern *M-L...L-M* spin exchanges.

As a representative example capturing the essence of spin exchange interactions, let us examine those of $LiCuVO_4$ [24-26] in which the $CuO_2$ ribbon chains, made up of edge-sharing $CuO_4$ square planes running along the b-direction are interconnected along the a-direction by sharing corners with $VO_4$ tetrahedra. This is shown in **Fig. 13**. In $LiCuVO_4$ the $Cu^{2+}$ (S = 1/2, $d^9$) ions are magnetic, but the $V^{5+}$ ($d^0$) ions are nonmagnetic. As for the spin exchange paths of $LiCuVO_4$, we consider the nearest neighbor (nn) and next-nearest-neighbor (nnn) intrachain spin exchanges, $J_{nn}$ and $J_{nnn}$, respectively, in each $CuO_2$ ribbon chain as well as the interchain spin exchange $J_a$ along the a-direction (**Fig. 13**).

The magnetic orbital of the $Cu^{2+}$ (S = 1/2, $d^9$) ion is given by the $x^2$-$y^2$ $\sigma$-antibonding orbital contained in the $CuO_4$ square plane (**Fig. 14a**), in which the Cu 3d $x^2$-$y^2$ orbital is combined out-of-phase with the 2p orbitals of the four surrounding O ligands. As already emphasized,[1,2] it is not the "head" part (the Cu 3d $x^2$-$y^2$ orbital) but the "tail" part (the O 2p orbitals) of the magnetic orbital that controls the magnitudes and signs of these spin exchange interactions. Let us first consider the Cu-O-Cu exchange $J_{nn}$. When the $x^2$-$y^2$ magnetic orbitals $\phi_1$ and $\phi_2$ of the two spin sites are brought together to form the Cu-O-Cu bridges, the O 2p orbital tails at the bridging O atoms make a nearly orthogonal arrangement (**Fig. 14b**). Thus, the overlap integral $\langle \phi_1 | \phi_2 \rangle$ between the two magnetic orbitals is almost zero, which leads to $J_{AF} \approx 0$. In contrast, the overlap density $\phi_1\phi_2$ of the magnetic orbitals is substantial, which leads to nonzero $J_F$. As a consequence, the $J_{nn}$ exchange becomes FM.[25,26]

For the intra-chain Cu-O…O-Cu exchange $J_{nnn}$ (**Fig. 14c**), the O 2p orbital tails of the magnetic orbitals $\phi_1$ and $\phi_2$ at the terminal O atoms are well separated by the O…O contacts. Thus, the overlap density $\phi_1\phi_2$ of the magnetic orbitals is negligible leading to $J_F$ ≈ 0. However, the overlap integral $\langle \phi_1 | \phi_2 \rangle$ is nonzero because the O 2p tails of $\phi_1$ and $\phi_2$ overlap through the O…O contacts. This through-space interaction between $\phi_1$ and $\phi_2$ produces a large energy split $\Delta e$ between $\psi_+$ and $\psi_-$, which are in-phase and out-of-phase combinations of $\phi_1$ and $\phi_2$, respectively (**Fig. 15a**), thereby leading to nonzero $J_{AF}$. Consequently, the $J_{nnn}$ exchange becomes AFM.[25,26]

In the interchain spin exchange path $J_a$, the two $CuO_4$ square planes are corner-shared with $VO_4$ tetrahedra. In the Cu-O…$V^{5+}$…O-Cu exchange paths, the empty V 3d orbitals should interact in a bonding manner with the Cu $x^2$-$y^2$ orbitals. In the absence of the V 3d orbitals, the energy split $\Delta e$ between $\psi_+$ and $\psi_-$ arising from the through-space interaction between $\phi_1$ and $\phi_2$ would be substantial, as expected from the intrachain exchange $J_{nnn}$, so that one might expect a strong AFM exchange for the interchain exchange $J_a$. However, in the Cu-O…$V^{5+}$…O-Cu exchange paths, the bridging $VO_4$ units provides a through-bond interaction between the empty V $3d_\pi$ orbitals and the O 2p tails of the magnetic orbitals on the O…O contacts. By symmetry, this through-bond interaction is possible only with $\psi_-$ (**Fig. 15b,c**). The V $3d_\pi$ orbital being empty, the O 2p tails of $\psi_-$ on the O…O contacts interact in-phase with the empty V $3d_\pi$ orbital hence lowering the $\psi_-$ level, whereas $\psi_+$ is unaffected by the V $3d_\pi$ orbital, thereby reducing the energy split $\Delta e$ between $\psi_+$ and $\psi_-$ of the Cu-O…$V^{5+}$…O-Cu exchange paths and consequently weakening the interchain spin exchange $J_a$.[25,26] As a consequence, the magnetic properties are dominated by the one-dimensional character of the $CuO_2$ ribbon chain.

It is important to observe the corollary of the above observation for general *M-L…*$A^{y+}$*…L-M* spin exchange, where the cation $A^{y+}$ provides through-bond interactions. If the $\Delta e$ between $\psi_+$ and $\psi_-$ is negligible in terms of the through-space interaction, then the effect of the through-bond interaction would make $\Delta e$ large leading to a strong AFM interaction.[2]

When the $ML_n$ polyhedra containing *M* cations are condensed together by sharing a corner, an edge or a face, they give rise to *M-L-M* exchanges, which are the subject of the Goodenough rules.[22] When these polyhedra are not condensed, they give rise to *M-L...L-M* and *M-L…*$A^{y+}$*…L-M* exchanges,[1,2] where $A^{y+}$ (y = oxidation state) refers to the intervening $d^0$ metal cation. The importance of the latter spin exchanges, not covered by the Goodenough rules, was recognized[1,2] only after realizing that the magnetic orbitals of an M ion include both the *M* *d*-orbitals and the *L* *p*-orbitals of the $ML_n$ polyhedron, and that the *L* *p*-orbital tails of the magnetic orbitals control the magnitudes and signs of such spin exchange interactions.[1,2] Concerning the *M-L...L-M* exchanges, there are several important consequences of this observation:[1,2]

(a) The strength of a given *M-L...L-M* spin exchange is not determined by the shortness of the *M...M* distance, but rather by that of the *L...L* distance; it is strong when the *L...L* distance is in the vicinity of the van der Waals radii sum or shorter.[1]

(b) In a given magnetic system consisting of both *M-L-M* and *M-L...L-M* spin exchanges, the *M-L...L-M* spin exchanges are very often stronger than the *M-L-M* spin exchanges.

(c) The strength of an *M-L...L-M* spin exchange determined by through-space interaction between the L np tails on the *L...L* contact can be significantly modified when the *L...L* contact has a through-bond interaction with the intervening $d^0$ metal cation $A^{y+}$ (y = oxidation state)[25,27] or even the $p^0$ metal cation (e.g., $Cs^+$ as found for $Cs_2CuCl_4$ [28]). Such an *M-L...$A^{y+}$...L-M* spin exchange becomes strong if the corresponding *M-L...L-M* through-space exchange is weak, but becomes weaker if the corresponding *M-L...L-M* through-space exchange is strong. This is so because the empty $d_\pi$ orbital of $A^{y+}$ interacts only with the $\psi_-$ orbital of the *M-L...L-M* exchange. In general, the empty $d_\pi$ orbital has a much stronger through-bond effect than does the empty $p_\pi$ orbital.

## 5. Incorporating the effect of SOC indirectly into spin Hamiltonian

When a magnetic ion is present in molecules and solids to form a $ML_n$ polyhedron with surrounding ligands *L*, its orbital momentum $\vec{L}$ is mostly quenched with a small momentum $\delta\vec{L}$ remaining unquenched.[10] Exceptional cases occur when the $ML_n$ polyhedron has *n*-fold ($n \geq 3$) rotational symmetry so that it has doubly-degenerate *d*-states and when the d-electron count of $ML_n$ is such that a degenerate *d*-state is unevenly occupied. In this case, the orbital momentum $\vec{L}$ is not quenched so that the effect of the SOC, $\lambda\hat{S}\cdot\hat{L}$, becomes strong often leading to uniaxial magnetism (see Section 6). In this section, we consider the cases when the orbital quenching is not complete so a small orbital momentum $\delta\vec{L}$ remains at each magnetic ion. In the past this situation has been discussed on the basis of the effective spin approximation,[10,29] in which the need to explicitly describe the unquenched orbital momentum is circumvented by treating the system as a spin-only system. In this approximation the effect of SOC arising from $\delta\vec{L}$ is absorbed into the coefficient for certain terms made up of only spin operators. This approximation deals with both single-spin site and two-spin site problems. The former includes the single-ion anisotropy, while the latter include the asymmetric spin exchange and the Dzyaloshinskii-Moriya (DM) exchange.[30,31] The DM exchange is often referred to as antisymmetric exchange.

### 5.1. SOC effect on a single-spin site and spin-half misconception

For a magnetic ion with nondegenerate magnetic orbital (e.g., $Cu^{2+}$), the SOC Hamiltonian $\hat{H}_{SO} = \lambda\hat{S}\cdot\hat{L}$ is transformed into the zero-field spin Hamiltonian $\hat{H}_{zf}$ [10]

$$\begin{aligned}\hat{H}_{zf} &= D(\hat{S}_z^2 - \tfrac{1}{3}\hat{S}^2) + E(\hat{S}_x^2 - \hat{S}_y^2)\\ &= D(\hat{S}_z^2 - \tfrac{1}{3}\hat{S}^2) + \tfrac{1}{2}E(\hat{S}_+\hat{S}_+ + \hat{S}_-\hat{S}_-)\end{aligned} \qquad (16)$$

where the constants D and E originate from the SOC associated with the remnant orbital momentum $\delta\vec{L}$, that is,

$$D \propto \lambda^2(\delta L_\parallel - \delta L_\perp)$$

$$E \propto \lambda^2(\delta L_x - \delta L_y)$$

where $\delta L_\parallel$ and $\delta L_\perp$ are the the $\parallel$z- and $\perp$z-components of $\delta\vec{L}$, respectively, while $\delta L_x$ and $\delta L_y$ are the x- and y-components of $\delta L_\perp$, respectively.

For S > 1/2 ions, Eq. 16 predicts magnetic anisotropy. For instance, a S = 1 ion is described by three spin states, $|1,+1\rangle$, $|1,0\rangle$ and $|1,-1\rangle$. Thus,

$$D(\hat{S}_z^2 - \tfrac{1}{3}\hat{S}^2)|1,+1\rangle = D[\hat{S}_z^2 - \tfrac{1}{3}S(S+1)]|1,+1\rangle = +\tfrac{1}{3}D|1,+1\rangle$$

$$D(\hat{S}_z^2 - \tfrac{1}{3}\hat{S}^2)|1,0\rangle = D[\hat{S}_z^2 - \tfrac{1}{3}S(S+1)]|1,0\rangle = -\tfrac{2}{3}D|1,0\rangle$$

$$D(\hat{S}_z^2 - \tfrac{1}{3}\hat{S}^2)|1,-1\rangle = D[\hat{S}_z^2 - \tfrac{1}{3}S(S+1)]|1,-1\rangle = +\tfrac{1}{3}D|1,-1\rangle$$

and

$$E(\hat{S}_+\hat{S}_+ + \hat{S}_-\hat{S}_-)|1,+1\rangle = E|1,-1\rangle$$

$$E(\hat{S}_+\hat{S}_+ + \hat{S}_-\hat{S}_-)|1,0\rangle = 0$$

$$E(\hat{S}_+\hat{S}_+ + \hat{S}_-\hat{S}_-)|1,-1\rangle = E|1,+1\rangle$$

This shows that the $|1,\pm1\rangle$ states are separated in energy from the $|1,0\rangle$ state by $|D|$. In addition, the $|1,+1\rangle$ and $|1,-1\rangle$ states interact and become split in energy by $|E|$. Due to this energy split, the thermal populations of the three states differ, hence leading to magnetic anisotropy. A similar conclusion is reached for S > 1 ions. For example, a S = 3/2 ion is described by the four states, $\left|\tfrac{3}{2},+\tfrac{3}{2}\right\rangle$, $\left|\tfrac{3}{2},+\tfrac{1}{2}\right\rangle$, $\left|\tfrac{3}{2},-\tfrac{1}{2}\right\rangle$ and $\left|\tfrac{3}{2},-\tfrac{3}{2}\right\rangle$. Therefore,

$$D(\hat{S}_z^2 - \tfrac{1}{3}\hat{S}^2)\left|\tfrac{3}{2},+\tfrac{3}{2}\right\rangle = D[\hat{S}_z^2 - \tfrac{1}{3}S(S+1)]\left|\tfrac{3}{2},+\tfrac{3}{2}\right\rangle = D$$

$$D(\hat{S}_z^2 - \tfrac{1}{3}\hat{S}^2)\left|\tfrac{3}{2},+\tfrac{1}{2}\right\rangle = D[\hat{S}_z^2 - \tfrac{1}{3}S(S+1)]\left|\tfrac{3}{2},+\tfrac{1}{2}\right\rangle = 0$$

$$D(\hat{S}_z^2 - \tfrac{1}{3}\hat{S}^2)\left|\tfrac{3}{2},-\tfrac{1}{2}\right\rangle = D[\hat{S}_z^2 - \tfrac{1}{3}S(S+1)]\left|\tfrac{3}{2},+\tfrac{1}{2}\right\rangle = 0$$

$$D(\hat{S}_z^2 - \tfrac{1}{3}\hat{S}^2)\left|\tfrac{3}{2},-\tfrac{3}{2}\right\rangle = D[\hat{S}_z^2 - \tfrac{1}{3}S(S+1)]\left|\tfrac{3}{2},+\tfrac{3}{2}\right\rangle = D$$

and

$$E(\hat{S}_+\hat{S}_+ + \hat{S}_-\hat{S}_-)\left|\tfrac{3}{2},+\tfrac{3}{2}\right\rangle = E\left|\tfrac{3}{2},-\tfrac{1}{2}\right\rangle$$

$$E(\hat{S}_+\hat{S}_+ + \hat{S}_-\hat{S}_-)\left|\tfrac{3}{2},+\tfrac{1}{2}\right\rangle = E\left|\tfrac{3}{2},-\tfrac{3}{2}\right\rangle$$

$$E(\hat{S}_+\hat{S}_+ + \hat{S}_-\hat{S}_-)\left|\tfrac{3}{2},-\tfrac{1}{2}\right\rangle = E\left|\tfrac{3}{2},+\tfrac{3}{2}\right\rangle$$

$$E(\hat{S}_+\hat{S}_+ + \hat{S}_-\hat{S}_-)\left|\tfrac{3}{2},-\tfrac{3}{2}\right\rangle = E\left|\tfrac{3}{2},+\tfrac{1}{2}\right\rangle$$

Thus, the $\left|\tfrac{3}{2},\pm\tfrac{3}{2}\right\rangle$ states are separated in energy from the $\left|\tfrac{3}{2},\pm\tfrac{1}{2}\right\rangle$ states by $|D|$. Without loss of generality, it can be assumed that the $\left|\tfrac{3}{2},\pm\tfrac{3}{2}\right\rangle$ states lie higher than the $\left|\tfrac{3}{2},\pm\tfrac{1}{2}\right\rangle$

states. The $\left|\frac{3}{2},+\frac{3}{2}\right\rangle$ and $\left|\frac{3}{2},-\frac{1}{2}\right\rangle$ states interact with interaction energy E, and so are the states $\left|\frac{3}{2},-\frac{3}{2}\right\rangle$ and $\left|\frac{3}{2},+\frac{1}{2}\right\rangle$. Then, according to perturbation theory, the $\left|\frac{3}{2},\pm\frac{3}{2}\right\rangle$ states are raised in energy by E$^2$/|D|, and the $\left|\frac{3}{2},\pm\frac{1}{2}\right\rangle$ states are lowered in energy by E$^2$/|D|. Consequently, the $\left|\frac{3}{2},\pm\frac{3}{2}\right\rangle$ states become separated in energy from the $\left|\frac{3}{2},\pm\frac{1}{2}\right\rangle$ states by |D| + 2E$^2$/|D|.

The aforementioned energy split for S > 1/2 ions, and the associated magnetic anisotropy, is a consequence of SOC albeit indirectly through the constants D and E. Since the information about the orbital $\left|L,L_z\right\rangle$ of the magnetic ion is completely hidden in these constants, it is not possible to predict the preferred spin orientation of a S > 1/2 ion on the basis of Eq. 16, although one can infer that such an ion has magnetic anisotropy as described above.

A rather different situation occurs for a S = 1/2 ion, which is described by two spin states, $\left|\uparrow\right\rangle = \left|\frac{1}{2},+\frac{1}{2}\right\rangle$ and $\left|\downarrow\right\rangle = \left|\frac{1}{2},-\frac{1}{2}\right\rangle$. We note that

$$D(\hat{S}_z^2 - \tfrac{1}{3}\hat{S}^2)\left|\tfrac{1}{2},+\tfrac{1}{2}\right\rangle = D[\hat{S}_z^2 - \tfrac{1}{3}S(S+1)]\left|\tfrac{1}{2},+\tfrac{1}{2}\right\rangle = 0$$
$$D(\hat{S}_z^2 - \tfrac{1}{3}\hat{S}^2)\left|\tfrac{1}{2},-\tfrac{1}{2}\right\rangle = D[\hat{S}_z^2 - \tfrac{1}{3}S(S+1)]\left|\tfrac{1}{2},-\tfrac{1}{2}\right\rangle = 0$$

and

$$E(\hat{S}_+\hat{S}_+ + \hat{S}_-\hat{S}_-)\left|\tfrac{1}{2},+\tfrac{1}{2}\right\rangle = 0$$
$$E(\hat{S}_+\hat{S}_+ + \hat{S}_-\hat{S}_-)\left|\tfrac{1}{2},-\tfrac{1}{2}\right\rangle = 0$$

Consequently, the up-spin and down-spin states do not interact under $\hat{H}_{zf}$, so their degeneracy is not split. (This result obeys the Kramers degeneracy theorem,[32] which states that the degeneracy of an odd-spin system should not be split in the absence of an external magnetic field.) This is so even though the constants D and E are nonzero, that is, even though SOC effects are taken into consideration though indirectly. Thus, the thermal populations of the two states $|\uparrow\rangle$ and $|\downarrow\rangle$ are identical, hence leading to the conclusion that an S = 1/2 ion has no magnetic anisotropy that arise from SOC. This is the origin of the spin-half misconception.

Note that $\hat{H}_{SO} = \lambda\hat{S}\cdot\hat{L}$ and $\hat{H}_{zf}$ are local (i.e., single-spin site) operators, and do not describe interactions between different spin sites. The SOC-induced magnetic anisotropy for S > 1/2 ions is commonly referred to as the single-ion anisotropy, to which practitioners of spin Hamiltonian analysis have no objection. However, most of them deny strenuously that S = 1/2 ions have single-ion anisotropy and suggest the use of the term "magneto-crystalline anisotropy" to describe the experimentally observed magnetic anisotropy of S = 1/2 ions. In the vernacular this term is a red herring, because it means that the observed anisotropy is not caused by the single-spin site effect (i.e., SOC) but rather by nonlocal effects (i.e., anything other than SOC, e.g., asymmetric spin exchange and magnetic dipole-dipole interactions), just as Moriya and Yoshida argued for the S = 1/2 system $CuCl_2\cdot2H_2O$ more than six decades ago.[7] However, as recently shown [5,6,9] for various magnetic solids of S = 1/2 ions (see Section 7), the spin-half misconception is erroneous. Unfortunately, this misconception remains unabated because it is perpetuated in monographs and textbooks on magnetism.[8] In defense of the spin-half misconception, one might argue that the true magnetic energy states are not those generated by an electronic Hamiltonian, but those generated by a spin Hamiltonian. However, this argument is even more fallacious than the spin-half misconception, because it amounts to arguing that there exists no orbital momentum. The magnetic properties of a magnetic ion are ultimately related to its moment $\vec{\mu}$, which is the derivative of its total electronic energy with respect to an applied magnetic field (see Section 6).[10] The moment $\vec{\mu}$ consists of both orbital and spin components, i.e., $\vec{\mu} = \vec{\mu}_L + \vec{\mu}_S$, and these components are related to the orbital and spin momenta as $\vec{\mu}_L = -\mu_B \vec{L}$ and $\vec{\mu}_S = -2\mu_B \vec{S}$, where $\mu_B$ is the Bohr magneton. Consequently, the magnetic energy states become identical to those generated by a spin Hamiltonian, only if $\vec{L} = 0$, that is, only if the quenching of orbital momentum is complete. The latter condition is hardly met for all magnetic ions in molecules and solids. It is satisfied for all magnetic ions in a spin Hamiltonian analysis by definition. In short, S = 1/2 ions do possess single-ion anisotropy, but a spin Hamiltonian analysis predicts erroneously that they do not.

### 5.2. SOC effect on spin exchange: Mapping analysis for anisotropic spin exchange[33]

In some cases the spin exchange between two spin sites may not be isotropic. This is an indirect consequence of SOC because a spin at a given site has a preferred orientation due to SOC and because this orientation preference can influence the strength of the spin exchange. Given two spin sites, say, 1 and 2, one may take the z-axis along the exchange paths between 1 and 2. As already mentioned in Section 3, the anisotropic spin exchange interaction between two sites 1 and 2 is written as

$$\hat{H}_{spin} = J_x \hat{S}_{1x} \hat{S}_{2x} + J_y \hat{S}_{1y} \hat{S}_{2y} + J_z \hat{S}_{1z} \hat{S}_{2z}, \tag{17a}$$

To evaluate $J_x$, $J_y$ and $J_z$, we perform energy-mapping analysis by determining the energies of appropriate broken-symmetry spin states on the basis of DFT+U+SOC calculations. To determine the $J_x$ component, we consider the following four ordered spin states,

| State | Spin 1 | Spin 2 | Other spin site |
|-------|--------|--------|-----------------|
| 1 | (S, 0, 0) | (S, 0, 0) | Either (0, 0, S) or (0, 0, -S) |
| 2 | (S, 0, 0) | (-S, 0, 0) | according to the experimental (or a |
| 3 | (-S, 0, 0) | (S, 0, 0) | low-energy) spin state. Keep the |
| 4 | (-S, 0, 0) | (-S, 0, 0) | same for the four spin states |

Then, the energy difference, $E_1 + E_4 - E_2 - E_3$, of the four states is related to the spin exchange $J_x$ as,

$$J_x = \frac{E_1 + E_4 - E_2 - E_3}{4S^2}$$

Then, on the basis of DFT+U+SOC calculations for the four spin states, the value of $J_x$ is readily determined. The values of $J_y$ and $J_z$ are obtained in a similar manner. To obtain $J_y$, we do DFT+U+SOC calculations for the following states:

| State | Spin 1 | Spin 2 | Other spin site |
|-------|--------|--------|-----------------|
| 1 | (0, S, 0) | (0, S, 0) | Either (0, 0, S) or (0, 0, -S) |
| 2 | (0, S, 0) | (0, -S, 0) | according to the experimental (or a |
| 3 | (0, -S, 0) | (0, S, 0) | low-energy) spin state. Keep the |
| 4 | (0, -S, 0) | (0, -S, 0) | same for the four spin states |

Then, we find

$$J_y = \frac{E_1 + E_4 - E_2 - E_3}{4S^2}$$

To determine $J_z$, we perform DFT+U+SOC calculations for the following states:

| State | Spin 1 | Spin 2 | Other spin sites |
|-------|--------|--------|------------------|
| 1 | (0, 0, S) | (0, 0, S) | Either (S, 0, 0) or (-S, 0, 0) |
| 2 | (0, 0, S) | (0, 0, -S) | according to the experimental (or a |
| 3 | (0, 0, -S) | (0, 0, S) | low-energy) spin state. Keep the |
| 4 | (0, 0, -S) | (0, 0, -S) | same for the four spin states |

Then, we find

$$J_z = \frac{E_1 + E_4 - E_2 - E_3}{4S^2}$$

## 5.3. SOC effect on two adjacent spin sites

Another important consequence of SOC is the Dzyaloshinskii-Moriya (DM) interaction between two adjacent spin sites. Consider the SOC in a spin dimer made up of two spin sites 1 and 2, for which the SOC Hamiltonian is given by[10]

$$\hat{H}_{SO} = \lambda \hat{L} \cdot \hat{S} = \lambda(\hat{L}_1 + \hat{L}_2) \cdot (\hat{S}_1 + \hat{S}_2) \approx \lambda(\hat{L}_1 \cdot \hat{S}_1 + \hat{L}_2 \cdot \hat{S}_2). \tag{18}$$

where the last equality follows from the fact that the SOC is a local interaction. It is important to note that, although SOC describes a single-spin site interaction, the two spin sites can interact indirectly hence influencing their relative spin orientations.[2,31] As illustrated in **Fig. 16**, we suppose that an occupied orbital $\phi_i$ interacts with an unoccupied orbital $\phi_j$ at spin site 1 via SOC, and that the $\phi_i$ and $\phi_j$ of site 1 interact with an occupied orbital $\phi_k$ of site 2 via orbital interaction. The orbital mixing between $\phi_i$ and $\phi_k$ introduces the spin character of site 2 into $\phi_i$ of site 1, while that between $\phi_j$ and $\phi_k$ introduces the spin character of site 2 into $\phi_j$ of site 1. Namely,

$$\phi_i \rightarrow \phi_i' \approx (1 - \gamma^2)\phi_i + \gamma\phi_k$$
$$\phi_j \rightarrow \phi_j' \approx (1 - \gamma^2)\phi_j + \gamma\phi_k,$$

where $\gamma$ refers to a small mixing coefficient. Then, the SOC between such modified $\phi'_i$ and $\phi'_j$ at site 1 indirectly introduces the SOC-induced interaction between the spins at sites 1 and 2. For a spin dimer, there can be a number of interactions like the one depicted in **Fig. 16** at both spin sites, so summing up all such contributions gives rise to the DM interaction energy $E_{DM}$ between spin sites 1 and 2.

Suppose that $\delta\vec{L}_1$ and $\delta\vec{L}_2$ are the remnant orbital angular momenta at sites 1 and 2, respectively. Then, use of the $\hat{H}_{SO}$ (Eq. 18) as perturbation leads to the DM interaction energy $E_{DM}$,[10,31]

$$E_{DM} = [\lambda J_{12}(\delta\vec{L}_1 - \delta\vec{L}_2)] \cdot (\vec{S}_1 \times \vec{S}_2) \equiv \vec{D}_{12} \cdot (\vec{S}_1 \times \vec{S}_2)$$

In this expression, the DM vector $\vec{D}_{12}$ is related to the difference in the unquenched orbital angular momenta on the two magnetic sites 1 and 2, namely,

$$\vec{D}_{12} \equiv \lambda J_{12}(\delta\vec{L}_1 - \delta\vec{L}_2).$$

For a spin dimer with spin exchange $J_{12}$, the strength of its DM exchange $\vec{D}_{12}$ is discussed by considering the ratio $|D_{12}/J_{12}|$, which is often approximated by

$$|D_{12}/J_{12}| \approx \Delta g/g,$$

where $\Delta g$ is the contribution of the orbital moment to the g-factor g in the effective spin approximation. In general, the $\Delta g/g$ value is at most 0.1, so that the $|D_{12}/J_{12}|$ ratio is often expected to be 0.1 at most. However, it is important to recognize an implicit assumption behind this reasoning, namely, that the spin sites 1 and 2 have an identical chemical environment. When the two spin sites have different chemical environments, the $|D_{12}/J_{12}|$ ratio can be very large as found for a particular $Mn(2)^{3+}$-O-$Mn(3)^{4+}$ spin exchange path of CaMn$_7$O$_{12}$ (i.e., $|D_{12}/J_{12}| = 0.54$).[34] As depicted in **Fig. 16**, the magnitude of a DM vector $D_{12}$ is determined by the three matrix elements,

$$t_{SO} = \left\langle \phi_i \left| \hat{H}_{SO} \right| \phi_j \right\rangle, \quad t_{ik} = \left\langle \phi_i \left| \hat{H}^{eff} \right| \phi_k \right\rangle, \quad \text{and} \quad t_{jk} = \left\langle \phi_j \left| \hat{H}^{eff} \right| \phi_k \right\rangle.$$

When $t_{SO}$, $t_{ik}$ and $t_{jk}$ are all strong, the magnitude of the DM vector $D_{12}$ can be unusually large.[35]

### 5.4. Mapping analysis for the DM vector of an isolated spin dimer[2]

Let us consider how to determine the DM vector of an isolated spin dimer. So far, a spin dimer made up of spin sites 1 and 2 has been described by the spin Hamiltonian, $\hat{H}_{spin} = J_{12}\hat{S}_1 \cdot \hat{S}_2$, composed of only a Heisenberg spin exchange. This Hamiltonian leads to a collinear spin arrangement (either FM or AFM), as already mentioned. To allow for a canting of the spins $\vec{S}_1$ and $\vec{S}_2$ from the collinear arrangement (typically from the AFM arrangement), which is experimentally observed, it is necessary to include the DM exchange interaction $\vec{D}_{12} \cdot (\hat{S}_1 \times \hat{S}_2)$ into the spin Hamiltonian. That is,

$$\hat{H}_{spin} = J_{12}\hat{S}_1 \cdot \hat{S}_2 + \vec{D}_{12} \cdot (\hat{S}_1 \times \hat{S}_2). \tag{19}$$

The $\hat{S}_1 \times \hat{S}_2$ term, being proportional to $\sin\theta$, where $\theta$ is the angle between the two spin vectors $\vec{S}_1$ and $\vec{S}_2$, is nonzero only if the two spins are not collinear. Thus, the DM interaction $\vec{D}_{12} \cdot (\hat{S}_1 \times \hat{S}_2)$ induces spin canting. Even when a model Hamiltonian consists of only Heisenberg spin exchanges, a magnetic system with more than two spin sites can have a non-collinear spin arrangement so as to reduce the extent of spin frustration if there exists substantial spin frustration.

As discussed in Section 3; the spin exchange $J_{12}$ of Eq. 19 can be evaluated on the basis of energy-mapping analysis by considering two collinear spin states $|HS\rangle$ and $|BS\rangle$ (i.e., FM and AFM spin arrangements, respectively) because the DM exchange $\vec{D}_{12} \cdot (\hat{S}_1 \times \hat{S}_2)$ is zero for such collinear spin states. To evaluate the DM vector $\vec{D}_{12}$, we carry out energy-mapping analysis on the basis of DFT+U+SOC calculations. In terms of its Cartesian components, $\vec{D}_{12}$ is expressed as

$$\vec{D}_{12} = \hat{i}D_{12}^x + \hat{j}D_{12}^y + \hat{k}D_{12}^z$$

Therefore, the DM interaction energy $\vec{D}_{12} \cdot (\hat{S}_1 \times \hat{S}_2)$ is rewritten as

$$\vec{D}_{12} \cdot (\hat{S}_1 \times \hat{S}_2) = (\hat{i}D_{12}^x + \hat{j}D_{12}^y + \hat{k}D_{12}^z) \cdot \begin{pmatrix} \hat{i} & \hat{j} & \hat{k} \\ \hat{S}_1^x & \hat{S}_1^y & \hat{S}_1^z \\ \hat{S}_2^x & \hat{S}_2^y & \hat{S}_2^z \end{pmatrix}$$

$$= D_{12}^x(\hat{S}_1^y\hat{S}_2^z - \hat{S}_1^z\hat{S}_2^y) - D_{12}^y(\hat{S}_1^x\hat{S}_2^z - \hat{S}_1^z\hat{S}_2^x) + D_{12}^z(\hat{S}_1^x\hat{S}_2^y - \hat{S}_1^y\hat{S}_2^x) \qquad (20)$$

To determine the $D_{12}^z$ component, we consider the following two orthogonally ordered spin states,

| State | Spin 1 | Spin 2 |
|-------|--------|--------|
| 1 | (S, 0, 0) | (0, S, 0) |
| 2 | (S, 0, 0,) | (0, -S, 0) |

For these states, $\vec{S}_1 \cdot \vec{S}_2 = 0$ and $\left|\vec{S}_1 \times \vec{S}_2\right| = S^2$ so that, according to Eq. 20, the energies of the two states are given by

$$E_1 = S^2 D_{12}^z, \text{ and } E_2 = -S^2 D_{12}^z.$$

Consequently,

$$D_{12}^z = \frac{E_1 - E_2}{2S^2} \,.$$ (21a)

Thus, the $D_{12}^z$ is determined by evaluating the energies $E_1$ and $E_2$ on the basis of DFT+U+SOC calculations.

The $D_{12}^y$ and $D_{12}^x$ components are determined in a similar manner. Using the following two orthogonal spin states,

| State | Spin 1 | Spin 2 |
|-------|--------|--------|
| 3 | (S, 0, 0) | (0, 0, S) |
| 4 | (S, 0, 0,) | (0, 0, -S) |

the $D_{12}^y$ component is obtained as

$$D_{12}^y = \frac{E_3 - E_4}{2S^2}$$ (21b)

In terms of the following two orthogonal spin states,

| State | Spin 1 | Spin 2 |
|-------|--------|--------|
| 5 | (0, S, 0) | (0, 0, S) |
| 6 | (0, S, 0,) | (0, 0, -S) |

the $D_{12}^x$ term is given by

$$D_{12}^z = \frac{E_5 - E_6}{2S^2}$$ (21c)

## 5.4. Mapping analysis for the DM vectors using the four-state method for a general magnetic solid [4]

For a given pair of spins in a general magnetic solid, the $D_{12}^x$, $D_{12}^y$ and $D_{12}^z$ components can be similarly extracted by performing DFT+U+SOC calculations for four non-collinearly ordered spin states in which all spin exchange interactions associated with the spin sites 1 and 2 vanish.[4] In such a case the relative energies of the four states are related only to the energy differences in their DM interactions. To calculate the z-component of $D_{12}$, i.e., $D_{12}^z$, we carry out DFT+U+SOC calculations for the following four ordered spin states:

| State | Spin 1 | Spin 2 | Other spin sites |
|-------|--------|--------|------------------|
| 1 | (S, 0, 0) | (0, S, 0) | Either (0, 0, S) or (0, 0, -S) |
| 2 | (S, 0, 0) | (0, -S, 0) | according to the experimental (or a |
| 3 | (-S, 0, 0) | (0, S, 0) | low-energy) spin state. Keep the |
| 4 | (-S, 0, 0) | (0, -S, 0) | same for the four spin states |

Then, we obtain

$$D_{12}^z = \frac{E_1 + E_4 - E_2 - E_3}{4S^2} \qquad (22a)$$

To determine the y-component of $D_{12}$, i.e., $D_{12}^y$, we perform DFT+U+SOC calculations for the following four ordered spin states:

| State | Spin 1 | Spin 2 | Other spin site |
|-------|--------|--------|------------------|
| 1 | (S, 0, 0) | (0, 0, S) | Either (0, S, 0) or (0, -S, 0) |
| 2 | (S, 0, 0) | (0, 0, -S) | according to the experimental (or a |
| 3 | (-S, 0, 0) | (0, 0, S) | low-energy) spin state. Keep the |
| 4 | (-S, 0, 0) | (0, 0, -S) | same for the four spin states |

Then,

$$D_{12}^{y} = \frac{-E_1 - E_4 + E_2 + E_3}{4S^2} \qquad (22b)$$

To determine the x-component of $D_{12}$, i.e., $D_{12}^{x}$, we carry out DFT+U+SOC calculations for the following four ordered spin states:

| State | Spin 1 | Spin 2 | Other spin site |
|-------|--------|--------|-----------------|
| 1 | (0, S, 0) | (0, 0, S) | Either (S, 0, 0) or (-S, 0, 0) |
| 2 | (0, S, 0) | (0, 0, -S) | according to the experimental (or a |
| 3 | (0, -S, 0) | (0, 0, S) | low-energy) spin state. Keep the |
| 4 | (0, -S, 0) | (0, 0, -S) | same for the four spin states |

Then,

$$D_{12}^{x} = \frac{E_1 + E_4 - E_2 - E_3}{4S^2} \qquad (22c)$$

## 6. Uniaxial magnetism[10,36]

In classical mechanics, the magnetic moment $\vec{\mu}$ of a system refers to the change of its energy E with respect to the applied magnetic field $\vec{H}$,

$$\vec{\mu} = -\frac{\partial E}{\partial \vec{H}}. \qquad (23)$$

A uniaxial magnetic ion has a nonzero magnetic moment only in one direction in coordinate space, while an isotropic magnetic ion has a nonzero moment in all directions with equal magnitude. An anisotropic magnetic ion, lying between these two cases, has a moment with magnitude depending on the spin direction. When a transition-metal magnetic ion is located at a coordination site with 3-fold or higher rotational symmetry, its $d$-states have doubly-degenerate levels, namely,

{xz, yz} and {xy, x$^2$-y$^2$},

if the z-axis is taken along the rotational axis. In terms of the {$L_z$, $-L_z$} set of magnetic quantum numbers, the {xz, yz} and {xy, x$^2$-y$^2$} sets are equivalent to

{xz, yz} $\leftrightarrow$ {1, $-1$}

{xy, x$^2$-y$^2$} $\leftrightarrow$ {2, $-2$}

An uneven filling of such a degenerate level leading to configurations such as ($L_z$, $-L_z$)$^1$ and ($L_z$, $-L_z$)$^3$ generates an unquenched orbital angular momentum of magnitude L (in units of $\hbar$). Thus, an uneven filling of the {1, $-1$} set leads to L = 1, and that of the {2, $-2$} set to L = 2. Such an electron filling generates a Jahn-Teller (JT) instability, but the unquenched orbital momentum remains if the associated JT-distortion is prevented by steric congestion around the magnetic ions. The orbital momentum $\vec{L}$ couples with the spin momentum $\vec{S}$ by the SOC, $\lambda\vec{S}\cdot\vec{L}$, leading to the total angular momentum $\vec{J} = \vec{L} + \vec{S}$. The resulting total angular momentum states $\left|J, J_z\right\rangle$ are doublets specified by the two quantum numbers J and $J_z = \pm J$, i.e., {$\left|J, +J\right\rangle$, $\left|J, -J\right\rangle$}.[36] In identifying the ground doublet state, it is important to notice[10] that

$\lambda$ < 0 for an ion with more than half-filled d-shell
$\lambda$ > 0 for an ion with less than half-filled d-shell.

If $\lambda$ < 0, the lowest-energy doublet state of the $\lambda\vec{S}\cdot\vec{L}$ term results when $\vec{S}$ and $\vec{L}$ are in the same direction. If $\lambda$ > 0, however, it results when $\vec{S}$ and $\vec{L}$ have the opposite directions. Consequently, for a magnetic ion with L and S, the total angular quantum number J for the spin-orbit coupled ground state is given by

$$\text{Ground doublet}: \; J = \begin{cases} L+S & \text{if } \lambda < 0 \\ L-S & \text{if } \lambda > 0 \end{cases}.$$

For $\lambda < 0$, the energy of the J-state increases as J decreases. However, the opposite is the case for $\lambda > 0$.[36]

In quantum mechanical description, the moment is related to an energy split of a degenerate level by an applied magnetic field. The Zeeman interaction under magnetic field is given by[36]

$$\hat{H}_Z = \mu_B (\hat{L} + 2\hat{S}) \cdot \vec{H} \qquad (24)$$

If we take the z-axis along the rotational axis responsible for the degeneracy of the doublet state $\{\left| J,+J \right\rangle, \left| J,-J \right\rangle\}$, the Zeeman interaction for the field along the z-direction, $H_\parallel$, is written as

$$\hat{H}_{Z\parallel} = \mu_B (\hat{L}_z + 2\hat{S}_z) H_\parallel. \qquad (25a)$$

This Hamiltonian always lifts the degeneracy of $\{\left| J,+J \right\rangle, \left| J,-J \right\rangle\}$, because

$$\left\langle J,+J \right| \hat{H}_\parallel \left| J,+J \right\rangle = (L+2S)\mu_B H_\parallel$$
$$\left\langle J,-J \right| \hat{H}_\parallel \left| J,-J \right\rangle = -\left\langle J,+J \right| \hat{H}_\parallel \left| J,+J \right\rangle$$
$$\left\langle J,+J \right| \hat{H}_\parallel \left| J,-J \right\rangle = 0$$

Therefore, the energy split $\Delta E_{J\parallel}$ is given by

$$\Delta E_{J\parallel} = 2(L+2S)\mu_B H_\parallel \qquad (25b)$$

and the associated g-factor $g_\parallel$ by

$$g_{\parallel} = \Delta E_{J\parallel} / \mu_B H_{\parallel} = 2(L + 2S)$$

The Zeeman interaction for the field perpendicular to the z-direction, $H_{\perp}$, is written as

$$\hat{H}_{Z\perp} = \mu_B[\tfrac{1}{2}(\hat{L}_+ + \hat{L}_-) + (\hat{S}_+ + \hat{S}_-)]H_{\perp}, \qquad (26a)$$

for which we find

$$\langle J,+J | \hat{H}_{\perp} | J,+J \rangle = \langle J,-J | \hat{H}_{\perp} | J,-J \rangle = 0$$
$$\langle J,+J | \hat{H}_{\perp} | J,-J \rangle = \mu_B \langle J,+J | \tfrac{1}{2}(\hat{L}_+ + \hat{L}_-) + (\hat{S}_+ + \hat{S}_-) | J,-J \rangle \equiv \mu_B H_{\perp} \delta$$

Then, the associated energy split $\Delta E_{J\perp}$ is given by

$$\Delta E_{J\perp} = 2\mu_B H_{\perp} | \delta |. \qquad (26b)$$

The $|J,+J\rangle$ and $|J,-J\rangle$ states differ in their $J_z$ values by 2J, so $\Delta E_{J\perp} = 0$ unless J = 1/2 because $|J,-J\rangle$ state cannot become $|J,+J\rangle$ by the ladder operator $\hat{L}_+$ or $\hat{S}_+$ in such a case. Thus, for magnetic ions with unquenched orbital momentum $\vec{L}$, we find uniaxial magnetism if J > 1/2.[36]

It should be noted that a spin Hamiltonian does not allow one to predict whether or not a given magnetic ion in molecules and solids will exhibit uniaxial magnetism because it cannot describe SOC, $\lambda \hat{S} \cdot \hat{L}$, explicitly due to the lack of the orbital degree of freedom. Nevertheless, once a magnetic system is known to exhibit uniaxial magnetism, one might use an Ising spin Hamiltonian (Section 3) to discuss its magnetic property.

## 7. Describing SOC effects with both orbital and spin degrees of freedom: Magnetic anisotropy[5]

In this section we probe the effect of SOC by explicitly considering the orbital and spin degrees of freedom. This enables one to quantitatively determine the preferred spin orientation of a magnetic ion M with any spin (i.e., S = 1/2 – 5/2) by performing DFT+U+SOC calculations and qualitatively predict it on the basis of analyzing the HOMO-LUMO interactions of the $ML_n$ polyhedron induced by SOC, $\lambda \hat{S} \cdot \hat{L}$. For this purpose, the states of a magnetic ion are described by $\left| L, L_z \right\rangle \left| S, S_z \right\rangle$ instead of approximating it with $\left| S, S_z \right\rangle$. If a coordinate (x′, y′, z′) is employed for the spin $\hat{S}$, and (x, y, z) for the orbital $\hat{L}$, the z′ direction is the preferred spin orientation by convention. The latter is specified with respect to the (x, y, z) coordinate by defining the polar angles θ and φ as depicted in **Fig. 17**. In evaluating whether or not the SOC-induced interactions between different electronic states vanish, one needs to recall that the orbital states $\left| L, L_z \right\rangle$ are orthonormal, and so are the spin states $\left| S, S_{z'} \right\rangle$. That is,

$$\left\langle L, L_z \middle| L, L_z' \right\rangle = \begin{cases} 1, & \text{if } L_z = L_z' \\ 0, & \text{otherwise} \end{cases}$$

$$\left\langle S, S_z \middle| S, S_z' \right\rangle = \begin{cases} 1, & \text{if } S_z = S_z' \\ 0, & \text{otherwise} \end{cases}$$

### 7.1. Selection rules for preferred spin-orientation

Using the (x, y, z) and (x′, y′, z′) coordinates for $\hat{L}$ and $\hat{S}$, respectively, the SOC Hamiltonian $\hat{H} = \lambda \hat{S} \cdot \hat{L}$ is rewritten as $\hat{H} = \hat{H}_{SO}^0 + \hat{H}_{SO}'$,[2,10,37,38] where

$$\hat{H}_{SO}^0 = \lambda \hat{S}_{z'} \left( \hat{L}_z \cos\theta + \frac{1}{2} \hat{L}_+ e^{-i\phi} \sin\theta + \frac{1}{2} \hat{L}_- e^{+i\phi} \sin\theta \right) \tag{27a}$$

$$= \lambda \hat{S}_{z'} (\hat{L}_z \cos\theta + \hat{L}_x \sin\theta \cos\phi + \hat{L}_y \sin\theta \sin\phi) \,. \tag{27b}$$

$$\hat{H}'_{SO} = \frac{\lambda}{2}(\hat{S}_{+'} + \hat{S}_{-'})\left(-\hat{L}_z \sin\theta + \hat{L}_x \cos\theta\cos\phi + \hat{L}_y \cos\theta\sin\phi\right) \qquad (28)$$

We now consider if the preferred spin orientation is parallel to the local z-direction (∥z) (of the *ML*$_n$ under consideration) or perpendicular to it (⊥z). The SOC-induced interaction between two d-states, $\psi_i$ and $\psi_j$, involves the interaction energy $\left\langle\psi_i\middle|\hat{H}_{SO}\middle|\psi_j\right\rangle$. For our discussion, it is necessary to know whether this integral is zero or not. Since the angular part of a *d*- or *p*-orbital is expressed in terms of products $\left|L, L_z\right\rangle\left|S, S_{z'}\right\rangle$, the evaluation of $\left\langle\psi_i\middle|\hat{H}_{SO}\middle|\psi_j\right\rangle$ involves the spin integrals

$$\left\langle S, S'_{z'}\middle|\hat{S}_{z'}\middle|S, S_{z'}\right\rangle \text{ and } \left\langle S, S'_{z'}\middle|\hat{S}_{\pm'}\middle|S, S_{z'}\right\rangle$$

as well as the orbital integrals

$$\left\langle L, L'_z\middle|\hat{L}_z\middle|L, L_z\right\rangle \text{ and } \left\langle L, L'_z\middle|\hat{L}_{\pm}\middle|L, L_z\right\rangle.$$

The SOC Hamiltonian $\hat{H}^0_{SO}$ allows interactions only between identical spin states, because $\left\langle\uparrow\middle|\hat{S}_{z'}\middle|\uparrow\right\rangle$ and $\left\langle\downarrow\middle|\hat{S}_{z'}\middle|\downarrow\right\rangle$ are nonzero. For two states, $\psi_i$ and $\psi_j$, of identical spin, we consider the cases when $|\Delta L_z| = 0$ or 1. Then, we find

$$\left\langle\psi_i\middle|\hat{H}^0_{SO}\middle|\psi_j\right\rangle \propto \begin{cases} \cos\theta, & \text{if } |\Delta L_z| = 0 \\ \sin\theta, & \text{if } |\Delta L_z| = 1 \end{cases}. \qquad (29a)$$

For the $|\Delta L_z| = 0$ case, $\left\langle\psi_i\middle|\hat{H}^0_{SO}\middle|\psi_j\right\rangle$ is maximum at θ = 0°, i.e., when the spin has the ∥z orientation. For the $|\Delta L_z| = 1$ case, $\left\langle\psi_i\middle|\hat{H}^0_{SO}\middle|\psi_j\right\rangle$ becomes maximum at θ = 90°, i.e., when the spin has the ⊥z orientation. Under SOC $\psi_i$ and $\psi_j$ do not interact when $|\Delta L_z| > 1$, because $\left\langle\psi_i\middle|\hat{H}^0_{SO}\middle|\psi_j\right\rangle = 0$ in such a case.

The total energy of $ML_n$ is lowered under SOC by the interactions of the filled $d$-states with the empty ones. Since the strength of SOC is very weak, these interactions can be described in terms of perturbation theory in which the SOC Hamiltonian is taken as perturbation with the split $d$-states of $ML_n$ as unperturbed states. Then, the most important interaction of the occupied $d$-states with the unoccupied $d$-states is the one between the HOMO and the LUMO (with energies $e_{HO}$ and $e_{LU}$, respectively), and the associated energy stabilization $\Delta E$ is given by[5]

$$\Delta E = \begin{cases} -\left| \left\langle HO \left| \hat{H}_{SO}^0 \right| LU \right\rangle \right|, & \text{if } e_{HO} = e_{LU} \\[2em] -\dfrac{\left| \left\langle HO \left| \hat{H}_{SO}^0 \right| LU \right\rangle \right|^2}{\left| e_{HO} - e_{LU} \right|}, & \text{if } e_{HO} < e_{LU} \end{cases} \tag{29b}$$

Thus, we obtain the predictions for the preferred spin orientation as summarized in **Table 4**. In general, the effect of a degenerate interaction is stronger than that of a nondegenerate interaction. A system with degenerate HOMO and LUMO has JT instability, and the degeneracy would be lifted if the associated JT-distortion were to take place.[39]

According to Eq. 29, the preferred spin orientation is either $\|z$ or $\perp z$. For the preferred spin orientation to lie in between the $\|z$ and $\perp z$ directions, therefore, there must be two "HOMO-LUMO" interactions that predict different spin orientations (one for $\|z$, and the other for $\perp z$). Such a situation occurs for $Na_2IrO_3$, as will be discussed below.

### 7.2. Degenerate perturbation and uniaxial magnetism

For a certain metal ion $M$, the electron configuration of $ML_n$ has unevenly-filled degenerate level. For example, the hexagonal perovskites $Ca_3CoMnO_6$ [40] consist of $CoMnO_6$ chains in which $CoO_6$ trigonal prisms containing high-spin $Co^{2+}$ ($S = 3/2$, $d^7$) ions alternate with $MnO_6$ octahedra containing high-spin $Mn^{4+}$ ($S = 3/2$, $d^3$) ions by sharing their triangular faces (**Fig. 18a**). The $d$-states of the high-spin $Co^{2+}$ ($S = 3/2$, $d^7$) ion in each $CoO_6$ trigonal prism (**Fig. 18b**) can be described by the electron configuration, $(z^2)^2 < (xy, x^2–y^2)^3 < (xz, yz)^2$, in the one-electron picture.[36,39] Thus, the spin-polarized $d$-states of the high-spin $Co^{2+}$ is written as,

$(z^2\uparrow)^1 < (xy\uparrow, x^2–y^2\uparrow)^2 < (xz\uparrow, yz\uparrow)^2 < (z^2\downarrow)^1 < (xy\downarrow, x^2–y^2\downarrow)^1 < (xz\downarrow, yz\downarrow)^0$.

Due to the half-filled configuration $(xy\downarrow, x^2–y^2\downarrow)^1$, the HOMO and LUMO are degenerate with $|\Delta L_z| = 0$, so the preferred spin orientation is $\parallel z$, i.e., along the three-fold rotational axis of the trigonal prism. Furthermore, the configuration $(xy\downarrow, x^2–y^2\downarrow)^1$ leads to an unquenched orbital momentum for $L = 2$. Since the $d$-shell of the high-spin $Co^{2+}$ ($d^7$, $S = 3/2$) ion is more than half filled, $\lambda < 0$, so that $J = L + S = 2 + 3/2 = 7/2$ for the ground doublet state. Since $J > 1/2$, this ion has uniaxial magnetism, that is, it has a nonzero magnetic moment $\vec{\mu}$ only along the 3-fold rotational axis of the $CoO_6$ trigonal prism.

Each high-spin $Fe^{2+}$ ($S = 2$, $d^6$) ion of $Fe[C(Si(CH_3)_3)_3]_2$ is located at a linear coordinate site (**Fig. 5d**),[36,41] so that its down-spin d-states are filled as depicted in **Fig. 19** leading to the configuration $(xy\downarrow, x^2-y^2\downarrow)^1$. Thus, with $L = 2$ and $S = 2$, the spin-orbit coupled ground doublet state is described by $J = L + S = 4$ with $J_z = \pm 4$. Since $J > 1/2$, this ion has uniaxial magnetism; it has a nonzero magnetic moment $\vec{\mu}$ only along the C-

Fe-C axis (i.e., along the $C_\infty$-rotational axis), and hence this $Fe^{2+}$ ion has uniaxial magnetism

We now examine the uniaxial magnetism that arises from metal ions at octahedral sites by considering the $FeO_6$ octahedra with high-spin $Fe^{2+}$ ($d^6$, S = 2) ions present in the oxide $BaFe_2(PO_4)_2$, the honeycomb layers of which are made up of edge-sharing $FeO_6$ octahedra. This oxide exhibits a uniaxial magnetism.[42] For our analysis of this observation, it is convenient to take the z-axis along one three-fold rotational axis of an $ML_6$ octahedron (**Fig. 6a**).[12] The high-spin $Fe^{2+}$ ion has the $(t_{2g})^4(e_g)^2$ configuration, the $(t_{2g})^4$ configuration of which can be described by $\Psi_{Fe,1}$ or $\Psi_{Fe,2}$ shown below

$$\Psi_{Fe,1} = (1a)^1(1e_x, 1e_y)^3 = (1a\uparrow)^1(1e_x\uparrow, 1e_y\uparrow)^2(1e_x\downarrow, 1e_y\downarrow)^1$$
$$\Psi_{Fe,2} = (1a)^2(1e_x, 1e_y)^2 = (1a\uparrow)^1(1e_x\uparrow, 1e_y\uparrow)^2(1a\downarrow)^1$$

The occupancy of the down-spin $d$-states for $\Psi_{Fe,1}$ and $\Psi_{Fe,2}$ are presented in **Fig. 20a** and **20b**, respectively. An energy-lowering through SOC is strong for $\Psi_{Fe,1}$ because it has an unevenly filled degenerate configuration $(1e_x\downarrow, 1e_y\downarrow)^1$, but not by $\Psi_{Fe,2}$ because the latter has an evenly filled degenerate configuration $(1e_x\downarrow, 1e_y\downarrow)^2$. According to **Table 3**, the down-spin configuration $(1e_x\downarrow, 1e_y\downarrow)^1$ of $\Psi_{Fe,1}$ is expressed as

$$(1e_x\downarrow, 1e_y\downarrow)^1 = \left(\sqrt{\tfrac{2}{3}}(xy\downarrow, x^2-y^2\downarrow)^1 - \sqrt{\tfrac{1}{3}}(xz\downarrow, yz\downarrow)^1\right). \tag{30}$$

The orbital-unquenched state $(xy\downarrow, x^2-y^2\downarrow)^1$ leads to L = 2, but the state $(xz\downarrow, yz\downarrow)^1$ to L = 1. The SOC constant $\lambda < 0$ for the $\Psi_{Fe,1}$ configuration of $Fe^{2+}$ (S = 2, $d^6$) so that the ground doublet is J = L + S = 4 from the component $(xy\downarrow, x^2-y^2\downarrow)^1$ (L = 2), and J = 3 from $(xz\downarrow, yz\downarrow)^1$ (L = 1). In terms of the notation $\{J_z, -J_z\}$ representing a spin-orbit coupled doublet set, the doublet $\{4,-4\}$ is more stable than $\{3,-3\}$ because $\lambda < 0$, so the $(1e_x, 1e_y)^3$ configuration of $Fe^{2+}$ is expressed as

$$Fe^{2+} : (1e_x, 1e_y)^3 = (1e_x \uparrow, 1e_y \uparrow)^2 (1e_x \downarrow, 1e_y \downarrow)^1 \equiv \{4,-4\}_J^2 \{3,-3\}_J^1$$

With $J = 3$ for the singly-filled doublet, uniaxial magnetism is predicted for the high-spin $Fe^{2+}$ ion at an octahedral site with ∥z spin orientation. In support of this analysis, DFT calculations show the orbital moment of the $Fe^{2+}$ ion to be ~1 $\mu_B$ (i.e., $L \approx 1$).[43] Note that the $\Psi_{Fe,2}$ configuration (**Fig. 20b**) leads to $|\Delta L_z| = 1$ and hence the preference for the ⊥z spin orientation.

### 7.3. Nondegenerate perturbation and weak magnetic anisotropy

We now examine the preferred spin orientations of magnetic ions with nondegenerate HOMO and LUMO. The layered compound $SrFeO_2$ consists of $FeO_2$ layers made up of corner-sharing $FeO_4$ square planes containing high-spin $Fe^{2+}$ ($d^6$, S = 2) ions.[44] Corner-sharing $FeO_4$ square planes are also found in $Sr_3Fe_2O_5$, in which they form two-leg ladder chains.[45] The d-states of a $FeO_4$ square plane are split as in **Fig. 5c**,[46,47] so that the down-spin d-states have only the $3z^2-r^2\downarrow$ level filled, with the empty $\{xz\downarrow, yz\downarrow\}$ set lying immediately above (**Fig. 9**). Thus, between these HOMO and LUMO, with $|\Delta L_z| = 1$ so the preferred spin direction is ⊥z, i.e., parallel to the $FeO_4$ plane.[46,47]

A regular $MnO_6$ octahedron containing a high-spin $Mn^{3+}$ ($d^4$, S = 2) ion has JT instability and hence adopts an axially-elongated $MnO_6$ octahedron (**Fig. 5b**). Such JT-distorted $MnO_6$ octahedra are found in $TbMnO_3$ [48] and $Ag_2MnO_2$.[49,50] The neutron diffraction studies show that the spins of the $Mn^{3+}$ ions are aligned along the elongated

Mn-O bonds.[48,50] With four unpaired electrons to fill the split $d$-states, the LUMO is the $x^2-y^2\uparrow$ and the HOMO is the $3z^2-r^2\uparrow$ (**Fig. 21**). Between these two states, $|\Delta L_z| = 2$ so that they do not interact under SOC. The closest-lying filled $d$-state that can interact with the LUMO is the $xy\uparrow$. Now, $|\Delta L_z| = 0$ between the $x^2-y^2\uparrow$ and $xy\uparrow$ states, the preferred spin orientation is $\|z$, i.e., parallel to the elongated Mn-O bonds.[50,51]

The $NiO_6$ trigonal prisms containing $Ni^{2+}$ ($d^8$, S = 1) ions are found in the $NiPtO_6$ chains of $Sr_3NiPtO_6$,[52] which is isostructural with $Ca_3CoMnO_6$. Each $NiPtO_6$ chain consists of face-sharing $NiO_6$ trigonal prisms and $PtO_6$ octahedra. The $Pt^{4+}$ ($d^6$, S = 0) ions are nonmagnetic. As depicted in **Fig. 22** for the down-spin $d$-states of $Ni^{2+}$ ($d^8$, S = 1), $|\Delta L_z| = 1$ between the HOMO and LUMO. Consequently, the preferred spin orientation of the $Ni^{2+}$ ($d^8$, S = 1) ion is $\perp z$, i.e., perpendicular to the $NiPtO_6$ chain. This in agreement with DFT calculations.[6]

### 7.4. Magnetic anisotropy of S = 1/2 systems and spin-half misconception

In this section we examine the experimentally observed magnetic anisotropies of various S = 1/2 ions $M$. These observations are correctly reproduced by DFT+U+SOC calculations and also correctly explained by the SOC-induced HOMO-LUMO interactions of their $ML_n$ polyhedra. The experimental and theoretical evidence against the spin-half misconception is overwhelming to say the least.

First, we consider the magnetic ions with S = 1/2 in which the HOMO and LUMO of the crystal-field $d$-states are not degenerate. An axially-elongated $IrO_6$ octahedra containing low-spin $Ir^{4+}$ ($d^5$, S = 1/2) ions are found in the layered compound $Sr_2IrO_4$, in which the corner-sharing of the $IrO_6$ octahedra using the equatorial oxygen atoms forms the $IrO_4$ layers with the elongated Ir-O bonds perpendicular to the layer.[53-55] The neutron diffraction studies of $Sr_2IrO_4$ show that the $Ir^{4+}$ spins are parallel to the $IrO_4$ layer.[54,55] With the z-axis chosen along the elongated Ir-O bond, the $t_{2g}$ level of the $IrO_6$ octahedron is split into {xz, yz} < xy. With five *d*-electrons to fill the three levels, the down-spin states xz↓ and yz↓ are filled while the xy↓ state is empty, as depicted in **Fig. 23a**. Consequently, $|\Delta L_z| = 1$ between the HOMO and LUMO, so that the preferred spin orientation is ⊥z. This is in agreement with experiment and DFT calculations (See Section 8.1 for further discussion).[6,56]

$Na_2IrO_3$ consists of honeycomb layers made up of edge-sharing $IrO_6$ octahedra,[57,58] which are substantially compressed along the direction perpendicular to the layer (lying in the ab-plane), i.e., the c*-direction. Strictly speaking, each $IrO_6$ octahedron of $Na_2IrO_3$ has no 3-fold-rotational symmetry but has a pseudo 3-fold rotational axis along the c*-direction, which we take as the local z-axis. As for the preferred spin orientation of the $Ir^{4+}$ ions of $Na_2IrO_3$, experimental studies have not been unequivocal, nor have been DFT studies, but it has become clear that the preferred spin orientation has components along the c*- and a-directions (namely, ‖z and ⊥z components).[6,59,60] Due to the compression of the $IrO_6$ octahedron along this axis, its $t_{2g}$ state is split into $1a < (1e_x, 1e_y)$, where $1e_x$ and $1e_y$ are approximately degenerate, so that the down-spin d-states would be occupied as depicted in **Fig. 23b**. For the $Ir^{4+}$ ion of $Na_2IrO_3$, therefore, the HOMO and LUMO occur from the down-spin electron configuration close to $(1a\downarrow)^1(1e_x\downarrow, 1e_y\downarrow)^1$, so the preferred spin orientation would be the ‖z direction (namely, the c*-direction) because $\left|\Delta L_z\right| = 0$. The electron configuration $(1a\downarrow)^1(1e_x\downarrow, 1e_y\downarrow)^1$, deduced from an isolated $IrO_6$ octahedron, explains the c*-axis component, but cannot explain the presence of the a-axis component in the observed spin moment.[59-61] The perturbation theory analysis requires the split d-states of an $IrO_6$ octahedron present in $Na_2IrO_3$, not an isolated $IrO_6$ octahedron. The former have the effect of the intersite interactions, but the latter do not. Analysis of the intersite interaction showed [6] that they effectively reduce the energy split between $1a\downarrow$ and $(1e_x\downarrow, 1e_y\downarrow)$, so the $(1a\downarrow)^0(1e_x\downarrow, 1e_y\downarrow)^2$ configuration also participates substantially in controlling the spin orientation thereby giving rise to the a-axis component (See Section 8.1).

$CuCl_2\cdot2H_2O$ is a molecular crystal made up of $CuCl_2(OH_2)_2$ complexes containing $Cu^{2+}$ ($d^9$, S = 1/2) ions, in which the linear O-Cu-O unit is perpendicular to the linear Cl-Cu-Cl unit (**Fig. 24a**).[62] The spins of the $Cu^{2+}$ ions are aligned along the Cu-O direction,[63] namely, the $Cu^{2+}$ ions have easy-plane anisotropy. The split down-spin $d$-states of $CuCl_2\cdot2H_2O$ show that the LUMO, $x^2-y^2\downarrow$ has the smallest energy gap with the HOMO, $xz\downarrow$ (**Fig. 24b**).[9] Since $|\Delta L_z| = 1$, the preferred spin orientation is $\perp z$. To see if the spin prefers the x- or y-direction in the xy-plane, we use Eq. 27b. The matrix elements $\langle\psi_i|\hat{L}_\mu|\psi_j\rangle$ of the angular momentum operators $\hat{L}_\mu(\mu = x, y, z)$ are nonzero only for the following $\{\psi_i, \psi_j\}$ sets (see **Table 2**):[9]

For $\hat{L}_z$:     $\{xz, yz\}$,    $\{xy, x^2-y^2\}$
For $\hat{L}_x$:     $\{yz, 3z^2-r^2\}$,  $\{yz, x^2-y^2\}$,  $\{xz, xy\}$
For $\hat{L}_y$:     $\{xz, 3z^2-r^2\}$,  $\{xz, x^2-y^2\}$,  $\{yz, xy\}$

The only nonzero interaction between the LUMO $x^2-y^2\downarrow$ and the HOMO $xz\downarrow$ under SOC is the term $\langle x^2-y^2|\hat{L}_y|xz\rangle$ involving $\hat{L}_y$. Eq. 27b shows that this term comes with angular dependency of $\sin\theta\sin\phi$, which is maximized when $\theta = 90°$ and $\phi = 90°$. Thus, the preferred spin orientation of $CuCl_2(OH_2)_2$ is along the y-direction, namely, along the Cu-O bonds.[9]

In $CuCl_2$,[64,65] $CuBr_2$ [66] and $LiCuVO_4$,[24] the square planar $CuL_4$ units (L = Cl, Br, O) share their opposite edges to form $CuL_2$ ribbon chains (**Fig. 25a**). The split d-states in the $CuL_2$ ribbon chains of $CuCl_2$, $CuBr_2$ and $LiCuVO_4$ can be deduced by examining their projected density of states (PDOS) plots. Analyses of these plots can be best described by the effective sequence of the down-spin d-states shown in Eq. 31a.[9]

$$(3z^2-r^2\downarrow)^1(xy\downarrow)^1(xz\downarrow, yz\downarrow)^2(x^2-y^2\downarrow)^0 \text{ for a } CuL_4 \text{ of a } CuL_2 \text{ ribbon chain} \quad (31a)$$

$$(3z^2-r^2\downarrow)^1(xz\downarrow, yz\downarrow)^2(xy\downarrow)^1(x^2-y^2\downarrow)^0 \text{ for an isolated } CuL_4 \text{ square plane} \quad (31b)$$

Consequently, the interaction of the LUMO $x^2$-$y^2\downarrow$ with the HOMO ($xz\downarrow$, $yz\downarrow$) will lead to the $\perp z$ spin orientation for the $Cu^{2+}$ ions of the $CuL_2$ ribbon chains.[9] This down-spin d-state sequence is different from the corresponding one expected for an isolated $CuL_4$ square plane (shown in Eq. 31b). This is due to the orbital interactions between adjacent $CuL_4$ square planes in the $CuL_2$ ribbon chain, in particular, the direct metal-metal interactions involving the xy orbitals through the shared edges between adjacent $CuL_4$ square planes.

Now we consider the magnetic ions with S = 1/2 whose HOMO and LUMO are degenerate. $Sr_3NiIrO_6$ [67] is isostructural with $Ca_3CoMnO_6$, and its $NiIrO_6$ chains are made up of face-sharing $IrO_6$ octahedra and $NiO_6$ trigonal prisms. Each $NiO_6$ trigonal prism has a $Ni^{2+}$ ($d^8$, S = 1) ion, and each $IrO_6$ octahedron a low-spin $Ir^{4+}$ ($d^5$, S = 1/2) ion. Magnetic susceptibility and magnetization measurements[68,69] indicate that $Sr_3NiIrO_6$ has uniaxial magnetism with the spins of both $Ni^{2+}$ and $Ir^{4+}$ ions aligned along the chain direction. Neutron diffraction measurements show that in each chain the spins of adjacent

$Ni^{2+}$ and $Ir^{4+}$ ions are antiferromagnetically coupled.[68] The low-spin $Ir^{4+}(d^5$, S = 1/2) ion

has the configuration $(t_{2g})^5$, which can be represented by $\Psi_{Ir,1}$ or $\Psi_{Ir,2}$

$$\Psi_{Ir,1} = (1a)^2 (1e_x, 1e_y)^3$$
$$\Psi_{Ir,2} = (1a)^1 (1e_x, 1e_y)^4$$

The occupancies of the down-spin $d$-states for $\Psi_{Ir,1}$ and $\Psi_{Ir,2}$ are given as depicted in **Fig.**

**26a** and **26b**, respectively. It is $\Psi_{Ir,1}$, not $\Psi_{Ir,2}$, that can lower energy strongly under SOC.

The down-spin part $(1e_x \downarrow, 1e_y \downarrow)^1$ of the configuration $(1e_x, 1e_y)^3$ in $\Psi_{Ir,1}$ can be rewritten as in Eq. 30 so that L = 2. For the low-spin $Ir^{4+}$, $\lambda < 0$, because the $t_{2g}$-shell is

more than half-filled.[10] With S = 1/2, we have J = L + S = 5/2 from $(xy, x^2-y^2)^3$, and 3/2

from $(xz, yz)^3$. Thus, the $(1e_x, 1e_y)^3$ configuration of $Ir^{4+}$ is expressed as

$$Ir^{4+} : (1e_x, 1e_y)^3 = (1e_x \uparrow, 1e_y \uparrow)^2 (1e_x \downarrow, 1e_y \downarrow)^1 \equiv \{5/2, -5/2\}^2 \{3/2, -3/2\}^1$$

The singly-filled doublet has J = 3/2, so uniaxial magnetism is predicted with the spin

orientation along the ∥z direction. This explains why the S = 1/2 ion $Ir^{4+}$ ion exhibits a

strong magnetic anisotropy with the preferred spin direction along the z-axis. In contrast

to the case of $Sr_3NiPtO_6$, the $Ni^{2+}$ ions of $Sr_3NiIrO_6$ have the ∥z spin orientation. This is

due to the combined effect of the uniaxial magnetism of the $Ir^{4+}$ ions and the strong AFM

spin exchange between adjacent $Ir^{4+}$ and $Ni^{2+}$ ions in each $NiIrO_6$ chain, which overrides

the weak preference for the ⊥z spin orientation for the $Ni^{2+}$ ion in an "isolated $NiO_6$"

trigonal prism (See Section 8.1 for further discussions).[5,6]

Let us consider the spin orientation of the S = 1/2 ions $V^{4+}$ ($d^1$) in the $VO_6$

octahedra of $R_2V_2O_7$ (R = rare earth),[70] in which each $VO_6$ octahedron is axially compressed along the direction of its local three-fold rotational axis (**Fig. 27a**) so that its $t_{2g}$ state is split into the 1a < 1e pattern (**Fig. 27b**). With the local z-axis along the three-fold rotational axis of $VO_6$, the HOMO is the 1a$\uparrow$ state, which is represented by $3z^2-r^2\uparrow$, which interacts with the LUMO 1e$\uparrow$ = (1e$_x\uparrow$, 1e$_y\uparrow$) states under SOC through their (xz$\uparrow$, yz$\uparrow$) components. Consequently, $|\Delta L_z| = 1$ and the preferred spin orientation would be $\perp z$. However, the observed spin orientation is $\|z$,[71] which has also been confirmed by DFT calculations.[72] This finding is explained if the $V^{4+}$ ion has some uniaxial magnetic character despite that the HOMO and LUMO are not degenerate. For the latter to be true, the true ground state of each $V^{4+}$ ion in $R_2V_2O_7$ should be a "contaminated state" 1a′, which has some contributions of the 1e and 2e character of its isolated $VO_6$ octahedron, namely,

$$\left|1a'\right\rangle \propto \left|1a\right\rangle + \gamma\left|1e\right\rangle + \delta\left|2e\right\rangle$$

where $\gamma$ and $\delta$ are small mixing coefficients. This is possible because each $VO_6$ octahedron present in $R_2V_2O_7$ has a lower symmetry than does an isolated $VO_6$ octahedron. The $VO_6$ octahedra are corner-shared to form a tetrahedral cluster (**Fig. 27c**), and such tetrahedral clusters further share their corners to form a pyroclore lattice (**Fig. 27d**). Indeed, the PDOS plots for the up-spin d-states of the $V^{4+}$ ions in $R_2V_2O_7$ show the presence of slight contributions of the 1e and 2e states to the occupied 1a state.[5,72]

As reviewed above, both experimental and theoretical studies reveal that S = 1/2 ions do have magnetic anisotropy induced by SOC. The spin-half misconception is in clear contradiction to these experimental and theoretical observations. Due to the simplification it introduces for doing complex calculations, spin Hamiltonian has been a practical tool of choice in doing physics on magnetism and will remain so for some time to come. Nevertheless, this success does not justify the perpetuation of the spin-half misconception. This failure of a spin Hamiltonian should be considered as a small price to pay for the enormous gain it provides.

### 7.5. Ligand-controlled spin orientation

For the $CuBr_4$ square planes of $CuBr_2$ ribbon chain,[66] the $CuBr_5$ square pyramids of $(C_5H_{12}N)CuBr_3$,[73,74] and the $CrI_6$ octahedra of the layered compound $CrI_3$,[75,76] the ligand $L$ is heavier than $M$, so the SOC between two d-states of $ML_n$ results more from the SOC-induced interactions between the $p$-orbitals of the ligands $L$ rather than from those between the $d$-orbitals of $M$. We clarify this point by considering a square planar $ML_4$ using the coordinate system of **Fig. 25a**. The metal and ligand contributions in the yz, xy and $x^2$–$y^2$ states of $ML_4$ are shown in **Fig. 25b-d**, respectively. The SOC-induced interaction between different $d$-states can occur by the SOC of $M$, and also by that of each ligand $L$. The interaction between the z and {x, y} orbitals at each $L$ has $|\Delta L_z| = 1$, leading to the $\perp z$ spin orientation. In contrast, the interaction between the x and y orbitals at each L has $|\Delta L_z| = 0$, leading to the $\parallel z$ spin orientation (**Table 2**). When the ligand $L$ is much heavier than the metal $M$, the SOC constant $\lambda$ of $L$ is greater than that of $M$. Furthermore, such ligands $L$ possess diffuse and high-lying p-orbitals, which makes the magnetic orbitals of $ML_n$ dominated by the ligand $p$-orbitals and also makes the $d$-states of $ML_n$ weakly split. This makes the SOC effect in $ML_n$ dominated by the ligands.

### 7.6. High-spin $d^5$ systems

High-spin $d^5$ transition-metal ions with S = 5/2 possess a small nonzero orbital momentum $\delta \vec{L}$ and exhibit weakly preferred spin orientations. For such a magnetic ion, the SOC-induced HOMO-LUMO interaction should be based on the $\hat{H}'_{SO}$ term (Eq. 28), because the HOMO and LUMO occur from different spin states. The comparison of Eq. 27b with Eq. 28 reveals that the predictions concerning the ∥z vs. ⊥z spin orientation from the term $\hat{H}'_{SO}$ are exactly opposite to those from the term $\hat{H}^0_{SO}$.

A similar situation occurs for a $d^3$ magnetic ion at octahedral sites, as found for the $Os^{5+}$ ions in $Ca_2ScOsO_6$ [77] and the $Ir^{6+}$ ions in $Sr_2CuIrO_6$,[78] because such an ion has the $(t_{2g})^3$ configuration and because the $t_{2g}$ states are well separated in energy from the $e_g$ states. Thus, the occupied up-spin $t_{2g}$ states, $t_{2g}{\uparrow}$, become the HOMO, and the unoccupied down-spin $t_{2g}$ states, $t_{2g}{\downarrow}$, the LUMO. It is known[79] that the orbital momentum of such a cation can be discussed by using the pseudo-orbital states $\left| L', L'_z \right\rangle$ with $L' = 1$ and $L'_z = 1, 0, -1$. To a first approximation, therefore, the orbital momentum of such a $d^3$ magnetic ion is zero. However, the quenching of the orbital momentum is not complete so that a $(t_{2g})^3$ ion has a small nonzero orbital momentum $\delta\vec{L}$. Thus the preferred spin orientation of $(t_{2g})^3$ ions is governed by the SOC-induced HOMO-LUMO interaction based on the $\hat{H}'_{SO}$ term (Eq. 28).[80]

### 8. Magnetic properties of 5d ion oxides[6]

The *d* orbitals of 5d ions are more diffuse than those of 3d ions, so that electron correlation is much weaker for 5d ions than for 3d ions. For a given $MO_n$ polyhedron, the *M* 3*d* and O 2*p* orbitals do not differ strongly in their contractedness so that the associated crystal-field splitting of an isolated $MO_n$ polyhedron is strong. However, the *M* 5*d* orbitals are much more diffuse than O 2*p* orbitals so that the 5*d*-state splitting of an isolated $MO_n$ polyhedron is weak. In addition, the interactions between adjacent metal ions $M$ through the $M$-O-$M$ bridges are stronger for 5d ions than for 3d ions. Thus, for 5d ion oxides, the relative ordering of their split $d$-states deduced from an isolated $MO_n$ polyhedron might change by the interactions between adjacent metal ions (i.e., the intersite interactions). Furthermore, each of the crystal-field split $d$-states can be split further by SOC,[81] and this effect is much stronger for 5d ion oxides than for 3d ion oxides because the strength of SOC is much stronger for 5d ions than for 3d ions. The weak electron correlation and strong SOC in 5d ion oxides have important consequences, as discussed below.

## 8.1. Spin-orbit Mott insulating state and Madelung potential

The combination of strong SOC and weak electron correlation creates a magnetic insulating state, as first reported for $Ba_2NaOsO_6$ containing $Os^{7+}$ ($d^1$) ions.[81] This phenomenon, quite common in 5d ion oxides, was considered as a consequence of strong spin-orbital entanglement,[82] and the resulting magnetic insulating state is described as a SOC-induced Mott insulating state[83] or spin-orbit Mott insulating state.[84] Both $Sr_3NiIrO_6$ and $Sr_2IrO_4$ are magnetic insulators, namely, they have a band gap at all temperature.[85-90] $Na_2IrO_3$ has been thought to be a magnetic insulator,[91,92] but a recent DFT study suggested that it might be a Slater insulator.[93] The latter refers to a system with a partially-filled bands and weak electron correlation that opens a band gap when it undergoes a metal-insulator transition at a temperature below which an AFM ordering sets in.[94] In addition to the local factors affecting electron localization such as the oxidation state and the SOC constant λ of a metal ion $M$, the extent of electron localization is influenced by the Madelung potential acting at the $M$, which is a non-local factor.[6] The Madelung potentials acting on the $Ir^{4+}$ sites less negative (i.e., less attractive) for of $Na_2IrO_3$ than for $Sr_3NiIrO_6$ and $Sr_2IrO_4$, namely, the 5d electrons of an $Ir^{4+}$ ion would be less strongly bound (i.e., less strongly localized) to the ion.[6]

## 8.2. Influence of intersite interactions on crystal field-split $d$-states[6]

In predicting the preferred spin orientations of magnetic ions $M$ in magnetic oxides on the basis of the SOC-induced HOMO-LUMO interactions, the split d-states of their local $MO_n$ polyhedra are needed. As pointed out above, for oxides of 5d ions, the relative ordering of their split $d$-states deduced from an isolated $MO_n$ polyhedron might change by the intersite interaction. In the following we examine how the intersite interactions affect the split 5$d$-states of the $Ir^{4+}$ ions in $Sr_3NiIrO_6$, $Sr_2IrO_4$ and $Na_2IrO_3$ and explore their consequences.

The ESR study of $Sr_2IrO_4$ showed [95] that the g-factors of the $Ir^{4+}$ ion along the ‖c and ⊥c directions are explained if the $t_{2g}$-states are split as xy < (xz, yz) rather than as (xz, yz) < xy (discussed in Section 7.4). This finding, puzzling from the viewpoints of the split $t_{2g}$ states of an isolated $IrO_6$ octahedron, reflects[6] that the split $d$-state patterns of $Sr_2IrO_4$ differ from those of an isolated $IrO_6$ octahedron due to the intersite interactions. In each $IrO_4$ layer of $Sr_2IrO_4$ the Ir-O-Ir linkages in the ab-plane are bent as shown in **Fig. 28a**. This bending of the Ir-O-Ir linkages does not weaken the π-antibonding interactions between adjacent xz/yz orbitals (**Fig. 28b**), but does weaken those between adjacent xy orbitals (**Fig. 28c**). Namely, the π-type interactions between adjacent xz/yz orbitals are stronger than those between adjacent xy orbitals. The split $d$-states of an $IrO_6$ octahedron embedded in $Sr_2IrO_4$ and hence having the intersite interactions can be approximated by those of a dimer made up of two adjacent corner-sharing $IrO_6$ octahedra. Then, the interactions between two adjacent $Ir^{4+}$ sites alter the crystal-field split $t_{2g}$ states as depicted in **Fig. 28d**, so that the HOMO has the xy character, and the LUMO the xz/yz character. This picture explains the PDOS plots of $Sr_2IrO_4$ shown in **Fig. 28e**, and predicts the ⊥c spin orientation as does the crystal-field split $t_{2g}$ states of an isolated $IrO_6$ octahedron (**Fig. 23a**). In addition, this explains why the ESR results[95] of $Sr_2IrO_4$ are explained by the *d*-state ordering xy < (xz, yz), despite that it consists of axially-elongated $IrO_6$ octahedra.

In $Na_2IrO_3$, edge-sharing $IrO_6$ octahedra form honeycomb layers (**Fig. 29a**), and such layers are stacked along the c-direction (**Fig. 29b**). DFT+U+SOC calculations reveal that the preferred spin orientation of the $Ir^{4+}$ ions in $Na_2IrO_3$ has both ∥c* and ∥a components.[6,96] To examine the cause for this observation, we consider how the intersite interaction affects relative ordering of the down-spin 1a and 1e states of an $Ir^{4+}$ ion (**Fig. 23b**). Consider a dimer made up of two adjacent $Ir^{4+}$ ions and recall that the *d*-orbital component of the 1a state is the $3z^2$-$r^2$ orbital, while those of the 1e state are the (xy, $x^2$-$y^2$) and (xz, yz) orbitals (**Table 3**). As depicted in **Fig. 29c**, the intersite interaction between the two 1a states leads to the $1a_+$ and $1a_-$ states, and that between the 1e states to the $1e_+$ and $1e_-$ states. The split between $1a_+$ and $1a_-$ states is weak because the lateral extension of the $3z^2$-$r^2$ orbitals within the plane of the honeycomb layer is small. In contrast, the split between the $1e_+$ and $1e_-$ states is large because the lateral extension of the (xy, $x^2$-$y^2$) orbitals is large and because so is that of the (xz, yz) orbitals. With four down-spin electrons in the dimer, the $1e_-$ states are empty while the remaining states are filled. The $|\Delta L_z| = 1$ interactions between the $1a_+/1a_-$ and $1e_-$ states predict the $\perp z$ spin orientation. The interactions between the $1e_+$ and $1e_-$ states give rise to the $|\Delta L_z| = 0$ interactions, between their (xz, yz) sets and between their (xy, $x^2$-$y^2$) sets, predicting the $\|z$ spin orientation. Consequently, if the $1a_+$ and $1a_-$ states are close in energy to the $1e_+$ states, then the preferred spin orientation of the $Ir^{4+}$ ion would be the ($\perp z + \|z$) direction. In essence, the $\|a$ component of the $Ir^{4+}$ spin orientation in $Na_2IrO_3$ is a consequence of the intersite interactions, because only the $\|c*$ direction is predicted in their absence.

The magnetic insulating state of $Sr_3NiIrO_6$ is reproduced by DFT+U+SOC calculations only when adjacent $Ni^{2+}$ and $Ir^{4+}$ spins have an AFM coupling in each $NiIrO_6$ chain.[6,97,98] It is known experimentally[85,86] that the preferred orientation of the $Ir^{4+}$ spins is the $\|c$-direction. DFT+U+SOC calculations showed that the preferred orientation of the $Ir^{4+}$ spins is the $\|c$-direction if the $Ni^{2+}$ and $Ir^{4+}$ spins have an AFM coupling,[6] but it is the $\perp c$-direction if they have an FM coupling.[6,99] In each $NiIrO_6$ chain the nearest-neighbor Ir…Ni distance is short due to the face-sharing between the $IrO_6$ and $NiO_6$ polyhedra so that the overlap between the Ir and Ni $3z^2$-$r^2$ orbitals can be strong. As illustrated in **Fig. 30a** and **30b**, the Ni $3z^2$-$r^2$ orbital is closer in energy to the Ir $3z^2$-$r^2$ orbital when adjacent $Ni^{2+}$ and $Ir^{4+}$ spins have an FM coupling than when they have an AFM coupling (see Section 2.2.2 and **Fig. 8**). The latter makes the interaction between the Ir and Ni $3z^2$-$r^2$ states stronger for the FM than for the AFM spin arrangement.[2,10,39b] As a consequence, the resulting antibonding state $(3z^2$-$r^2)_-$ is unoccupied for the FM spin arrangement, but it is occupied for the AFM spin arrangement (**Fig. 30a** and **30b**), as found by DFT+U calculations for $Sr_3NiIrO_6$;[6] the PDOS plots for the FM and AFM arrangements, presented in **Fig. 30c** and **30d**, respectively, reveal that the AFM arrangement is consistent with the local electron configuration $(1a\downarrow)^1(1e_x\downarrow, 1e_y\downarrow)^1$ (**Fig. 26a**), predicting the $\|z$ spin orientation, while the FM arrangement is consistent with the local configuration $(1a\downarrow)^0(1e_x\downarrow, 1e_y\downarrow)^1$ (**Fig. 26b**), predicting the $\perp z$ spin orientation.

## 8.3. Perturbation theory analysis of preferred spin orientation[6]

The energy stabilization $\Delta E$ associated with the SOC-induced interaction between the HOMO and the LUMO (with energies $e_{HO}$ and $e_{LU}$, respectively) is given by Eq. 29b. For the $Ir^{4+}$ (low-spin $d^5$) ion systems $Sr_3NiIrO_6$, $Sr_2IrO_4$ and $Na_2IrO_3$, the overall widths of the $t_{2g}$-block bandwidths are of the order of 2 eV (i.e., 1.7, 2.6 and 2.4 eV, respectively from our DFT+U calculations) and the HOMO-LUMO energy differences $\left|e_{HO} - e_{LU}\right|$ values are of the order of 0.2 eV (0.2, 0.2 and 0.3 eV, respectively.[6] The SOC constant $\lambda$ of $Ir^{4+}$ is of the order of 0.5 eV [8e] so that $\lambda^2$ is comparable in magnitude to $\left|e_{HO} - e_{LU}\right|$ for the case of $e_{HO} < e_{LU}$. In such a case, use of perturbation theory does not lead to an accurate estimation of $\Delta E$. However, this does not affect our qualitative predictions of the preferred spin orientations, because the latter do not require a quantitative evaluation of $\Delta E$.

## 8.4. LS vs jj coupling scheme of SOC[6]

The effects of SOC are discussed in terms of either the LS or the jj coupling scheme depending on the strength of SOC. In the LS (or Russel-Saunders) scheme the electron spin momenta are summed up to find the total spin momentum $\vec{S} = \sum \vec{s}_i$, and the orbital momenta of individual electrons to find the total orbital momentum $\vec{L} = \sum \vec{l}_i$.

Then, the SOC is included to couple $\vec{S}$ and $\vec{L}$ to obtain the total angular momentum $\vec{J}$, leading to the SOC Hamiltonian, $\hat{H}_{SO} = \lambda \vec{S} \cdot \vec{L}$. The LS-coupling scheme is typically employed for elements with weak SOC (e.g., 3d- and 4d-elements). In this scheme the crystal-field split $d$-states of a $MO_n$ polyhedron are closely related to the orbital states $\left| L, L_z \right\rangle$ of $M$ in the up-spin $\left| \uparrow \right\rangle$ or down-spin state $\left| \downarrow \right\rangle$ magnetic orbitals of $MO_n$. As found for $Sr_3NiIrO_6$, $Sr_2IrO_4$ and $Na_2IrO_3$ [6] and for $Ba_2NaOsO_6$,[81] our analyses based on the LS-coupling scheme explain the spin-orbit Mott insulating states of these 5d oxides as well as their observed magnetic anisotropies.

The jj-coupling scheme, appropriate for elements with strong SOC (e.g., 4f and 5f elements), has recently become popular in discussing the spin-orbit Mott insulating states of 5d oxides.[82] In this scheme, the spin and orbital momenta are added to obtain the total angular momentum $\vec{j}_i = \vec{l}_i + \vec{s}_i$ for each electron of a magnetic ion $M$, and the $\vec{j}_i$'s of the individual electrons are added to find the total angular momentum, $\vec{J} = \sum \vec{j}_i$, of $M$. In this approach, it is not readily obvious how to relate the $\vec{J}$ states to the crystal-field split $d$-states of $MO_n$ unless the corresponding analysis is done by using the LS-coupling scheme, because the crystal-field split d-states of $MO_n$ are determined by the interactions of the orbital states $\left| L, L_z \right\rangle$ of $M$ with the 2p orbitals of the surrounding O ligands and because the information about the orbital states $\left| L, L_z \right\rangle$ of $M$ is completely hidden in the jj-coupling scheme. As a consequence, use of the jj scheme makes it difficult to predict such fundamental magnetic properties as the preferred spin orientation and the uniaxial magnetism of a magnetic ion $M$. The latter are readily predicted by the LS coupling scheme. As found for the $Ir^{4+}$ ion of $Sr_3NiIrO_6$, the need to employ "J-states" in the LS scheme arises only when a magnetic ion has an unevenly-filled degenerate $d$-state, leading to an unquenched orbital momentum $\vec{L}$ that combines with $\vec{S}$ to form $\vec{J} = \vec{S} + \vec{L}$. In the LS scheme, use of J-states is inappropriate for $Sr_2IrO_4$ and $Na_2IrO_3$ because they possess no unquenched orbital momentum $\vec{L}$ to combine with $\vec{S}$.

Studies on $Sr_3NiIrO_6$, $Sr_2IrO_4$ and $Na_2IrO_3$ [6] and on $Ba_2NaOsO_6$ [81] strongly suggest that the magnetic properties of the 5d oxides are better explained by the LS scheme than by the jj scheme. The latter implies that the spin-orbital entanglement in 5d elements is not as strong as has been assumed.[82] These conclusions are consistent with the view that SOC for 5d elements lies in between the LS- and jj-coupling schemes, but is closer to the LS scheme.[100]

## 9. Concluding remarks

In this chapter we have reviewed how to think about magnetic properties of solid state materials from the perspectives of an electronic Hamiltonian. On the quantitative level, use of this Hamiltonian enables one

(a) to determine the relative stabilities of various spin arrangements on the basis of DFT+U or DFT+U+SOC calculations,

(b) to evaluate the spin exchange and DM exchange parameters that a spin Hamiltonian requires by performing energy-mapping analysis based on DFT+U or DFT+U+SOC calculations, and

(c) to characterize the magnetic anisotropy of a magnetic ion by performing DFT+U+SOC calculations.

On the qualitative level, use of an electronic Hamiltonian allows one

(a) to examine spin lattices in terms of M-L-M as well as M-L…L-M spin exchanges,

(b) to discuss how the strengths of M-L…L-M spin exchanges are modified by through-space and through-bond interactions, and

(c) to predict/rationalize the preferred spin orientation of a magnetic ion on the basis of its SOC-induced HOMO-LUMO interactions.

The qualitative concepts governing these structure-property correlations help one organize/think about known experimental/theoretical observations, design new experiments to do and new calculations to perform, and predict/rationalize the outcomes of the new studies.

In the past, a spin lattice required for spin Hamiltonian analysis used to be chosen by inspecting the pattern of magnetic ion arrangement and employing the Goodenough rules,[22] which cover only *M-L-M* spin exchanges. Use of Goodenough rules often led to spin lattices that are inconsistent with the electronic structures of the magnetic systems they are supposed to describe, to find that Goodenough rules are not adequate enough. The reason for this observation is that *M-L-M* spin exchanges are frequently much weaker than those spin exchanges not covered by Goodenough rules, namely, *M-L...L-M* and/or *M-L*…$A^{y+}$…*L-M* spin exchanges. This is understandable, because Goodenough rules were formulated in the mid 1950's, when the magnetic orbitals of *M* ions were regarded as their singly-occupied pure *d*-orbitals of *M*. The importance of *M-L...L-M* and/or *M-L*…$A^{y+}$…*L-M* spin exchanges were recognized only in the late 1990's and the early 2000's, when it was realized[1,2] that the strengths of spin exchanges are not governed by the metal *d*-orbital components, but by the ligand *p*-orbital components, of the magnetic orbitals of *ML*$_n$. Quantitative evaluations of *M-L-M*, *M-L...L-M* and *M-* $L{\dots}A^{y+}{\dots}L\text{-}M$ spin exchanges became possible by the energy-mapping analysis[1-4] based on DFT+U calculations developed in the early 2000's. This quantitative analysis helps one find, for any magnetic system, the spin lattice consistent with its electronic structure.

The spin-orbit Mott insulating states of the 5d oxides $Sr_3NiIrO_6$, $Sr_2IrO_4$ and $Na_2IrO_3$ as well as $Ba_2NaOsO_6$ are well explained by analyses based on the LS-coupling scheme of SOC. Furthermore, their observed magnetic anisotropies are better explained by the LS scheme rather than by the jj scheme. Consequently, the spin-orbital entanglement invoked for 5d elements is not as strong as has been put forward.[82] These observations are in support of the view that SOC for 5d elements lies in between the LS- and jj-coupling schemes, but is closer to the LS-coupling scheme.[100]

A magnetic ion has a preferred spin orientation because SOC induces interactions among its crystal-field split $d$-states and because the associated energy lowering depends on the spin orientation. The preferred spin orientation of a magnetic ion is readily predicted on the basis of the selection rule involving the SOC-induced HOMO-LUMO interaction. In the electronic structure description of a magnetic ion, each of its states has both orbital and spin components, that is, each state is represented by a set of orbital/spin states $\left|L,L_z\right\rangle\left|S,S_z\right\rangle$. The states of a magnetic ion are modified by SOC, $\lambda\hat{S}\cdot\hat{L}$, because it induces intermixing between them, but this intermixing takes place only in the orbital component $\left|L,L_z\right\rangle$ of each state. This explains why a magnetic ion has magnetic anisotropy regardless of whether its spin is 1/2 or not. A spin Hamiltonian analysis fails to explain this fundamental result because it represents each magnetic state in terms of only spin states $\left|S,S_z\right\rangle$. The effects of SOC, $\lambda\hat{S}\cdot\hat{L}$, can be included into a spin Hamiltonian only indirectly by using the zero-field Hamiltonian $\hat{H}_{zf}$ (Eq. 16). This Hamiltonian does not allow one to predict the preferred spin orientation for S > 1/2 ions, although it shows the presence of magnetic anisotropy arising from SOC for such ions in agreement with experiment. As for the S = 1/2 ions, however, this Hamiltonian is downright incorrect because it predicts the absence of magnetic anisotropy induced by SOC, $\lambda \hat{S} \cdot \hat{L}$, not to mention that it cannot predict their preferred spin orientation.

It is high time for the proponents of the spin-half misconception to recognize this shortcoming of a spin Hamiltonian analysis. Nevertheless, we are not unaware of the astute observation by Max Planck: "A new scientific truth does not triumph by convincing its opponents and making them see the light, but rather because its opponents eventually die and a new generation grows up that is familiar with it."[101] This observation is more explicitly paraphrased as "Death is an essential element in the progress of science, since it takes care of conservative scientists of a previous generation reluctant to let go of an old, fallacious theory and embrace a new and accurate one."[102] The debate on the spin-half misconception, which has just begun,[5,6,9] is certainly not as grand and epochal as that on the earth- vs. sun-centered model of the planetary motion, the single- vs. multi-galaxy universe, or the classical vs. quantum theory in the past, but unmistakable parallels exist between them. It is our hope that the readers of this chapter will have an open-minded view on magnetism and avoid falling into such a conceptual trap as the spin-half misconception.

**Acknowledgments**

This research used resources of the National Energy Research Scientific Computing Center, a DOE Office of Science User Facility supported by the Office of Science of the U.S. Department of Energy under Contract No. DE-AC02-05CH11231. Work at Fudan was supported by NSFC (11374056), the Special Funds for Major State Basic Research (2012CB921400, 2015CB921700), Program for Professor of Special Appointment (Eastern Scholar), Qing Nian Bo Jian Program, and Fok Ying Tung Education Foundation. MHW would like to thank Dr. Reinhard K. Kremer, Prof. Hyun-Joo Koo, Dr. Changhoon Lee and Elijah E. Gordon for their invaluable discussions over the years.

Table 1. Angular properties of atomic *p*- and *d*-orbitals

| x | $\left(\left|1,-1\right\rangle - \left|1,+1\right\rangle\right)/\sqrt{2}$ |
|---|---|
| y | $i\left(\left|1,-1\right\rangle + \left|1,+1\right\rangle\right)/\sqrt{2}$ |
| z | $\left|1,0\right\rangle$ |

| $3z^2-r^2$ | $\left|2,0\right\rangle$ |
|---|---|
| xz | $\left(\left|2,-1\right\rangle - \left|2,+1\right\rangle\right)/\sqrt{2}$ |
| yz | $i\left(\left|2,-1\right\rangle + \left|2,+1\right\rangle\right)/\sqrt{2}$ |
| xy | $i\left(\left|2,-2\right\rangle - \left|2,+2\right\rangle\right)/\sqrt{2}$ |
| $x^2-y^2$ | $\left(\left|2,-2\right\rangle + \left|2,+2\right\rangle\right)/\sqrt{2}$ |

Table 2. Nonzero integrals of the angular momentum operators, $\langle i|\hat{L}_x|j\rangle$, $\langle i|\hat{L}_y|j\rangle$ and $\langle i|\hat{L}_z|j\rangle$, where $(i, j = x, y, z)$ or $(i, j = 3z^2-r^2, xz, yz, x^2-y^2, xy)$.

| $\hat{L}_z$ | $\langle y|\hat{L}_z|x\rangle = i$ |
|---|---|
| $\hat{L}_x$ | $\langle z|\hat{L}_x|y\rangle = i$ |
| $\hat{L}_y$ | $\langle x|\hat{L}_y|z\rangle = i$ |

| | |
|---|---|
| $\hat{L}_z$ | $\langle xy|\hat{L}_z|x^2-y^2\rangle = 2i$ |
| | $\langle xz|\hat{L}_z|yz\rangle = -i$ |
| $\hat{L}_x$ | $\langle 3z^2-r^2|\hat{L}_x|yz\rangle = i\sqrt{3}$ |
| | $\langle x^2-y^2|\hat{L}_x|yz\rangle = i$ |
| | $\langle xz|\hat{L}_x|xy\rangle = i$ |
| $\hat{L}_y$ | $\langle 3z^2-r^2|\hat{L}_y|xz\rangle = -i\sqrt{3}$ |
| | $\langle x^2-y^2|\hat{L}_y|xz\rangle = i$ |
| | $\langle yz|\hat{L}_y|xy\rangle = -i$ |

Table 3. Orbital character of the *d*-states of an ML$_6$ octahedron in two different settings of the Cartesian coordinates

| z-axis direction | Along one M-L bond (Fig. 4a) | Along one C$_3$-rotational axis (Fig. 6a) |
|---|---|---|
| t$_{2g}$ | xy | $1a \equiv 3z^2 - r^2$ |
| | xz | $1e_x \equiv \sqrt{\frac{2}{3}}xy - \sqrt{\frac{1}{3}}xz$ |
| | yz | $1e_y \equiv \sqrt{\frac{2}{3}}(x^2 - y^2) - \sqrt{\frac{1}{3}}yz$ |
| e$_g$ | x$^2$–y$^2$ | $2e_x \equiv \sqrt{\frac{1}{3}}xy + \sqrt{\frac{2}{3}}xz$ |
| | 3z$^2$–r$^2$ | $2e_y \equiv \sqrt{\frac{1}{3}}(x^2 - y^2) + \sqrt{\frac{2}{3}}yz$ |

Table 4. The preferred spin orientations of magnetic ions predicted using the $|\Delta L_z|$ values associated with the SOC-induced HOMO-LUMO interactions

| Spin orientation | Requirement | Interactions between |
|:---:|:---:|:---:|
| $\parallel z$ | $|\Delta L_z| = 0$ | xz and yz <br> xy and $x^2-y^2$ <br> x and y |
| $\perp z$ | $|\Delta L_z| = 1$ | $\{3z^2-r^2\}$ and $\{xz, yz\}$ <br> $\{xz, yz\}$ and $\{xy, x^2-y^2\}$ <br> z and $\{x, y\}$ |

**Figure captions**

Figure 1.        Close-packed energy states of a magnetic system, which arise from weak interactions among the unpaired electrons of its magnetic ions.

Figure 2.        Examples of simple spin lattices: an isolated spin dimer and a uniform chain requiring one spin exchange constant, and an alternating chain and a two-leg ladder requiring two spin exchange constants.

Figure 3.        Minimum difference in the magnetic quantum numbers, $|\Delta L_z|$, between pairs of (a) $d$-orbitals and (b) $p$-orbitals.

Figure 4.        (a) An ideal $ML_6$ octahedron with the local z-axis taken along one M-O bond, i.e., one 4-fold rotational axis. (b) The orbital compositions of the $t_{2g}$ and $e_g$ states. (c) The $\pi$-antibonding in the xy, xz and yz components of the $t_{2g}$ state, and the $\sigma$-antibonding in the $x^2-y^2$ and the $3z^2-r^2$ components of the $e_g$ state.

Figure 5.        The split $d$-states of (a) an ideal $ML_6$ octahedron, (b) an axially-elongated $ML_6$ octahedron, (c) a square planar $ML_4$, and (d) a linear $ML_2$.

Figure 6.        (a) An ideal $ML_6$ octahedron with the local z-axis taken along one 3-fold rotational axis. (b) The orbital compositions of the $t_{2g}$ and $e_g$ states as listed in Table 3.

Figure 7.    The split of the up-spin and down-spin states by an on-site repulsion U. These states are degenerate in the non- spin-polarized description (left), but are split in the spin-polarized description (right).

Figure 8.    The orbital interactions between two equivalent spin sites for cases when they have (a) a FM arrangement and (b) an AFM arrangement.

Figure 9.    The simulation of the split $d$-states obtained from DFT+U calculations in terms of those obtained from an effective one-electron Hamiltonian for a high-spin (S = 2) $d^6$ ion at a square planar site forming a FeL$_4$ square plane.

Figure 10.    A spin dimer made up of two equivalent spin sites with an unpaired electron at each site. The unpaired electrons at the sites 1 and 2 are accommodated in the orbitals $\phi_1$ and $\phi_2$, respectively, and the spin exchange constant J describes the strength and sign of the interaction between the two unpaired electrons.

Figure 11.    The interaction between the magnetic orbitals $\phi_1$ and $\phi_2$ of a spin dimer leading to the bonding and antibonding molecular orbitals $\psi_1$ and $\psi_2$ of the dimer, respectively, which are split by the energy $\Delta e$.

Figure 12.    The occupation of the molecular orbitals $\psi_1$ and $\psi_2$ of the dimer with two electrons leading to the triplet configuration $\Psi_T$ as well as two singlet configurations $\Phi_1$ and $\Phi_2$.

Figure 13.    Two $CuO_2$ ribbon chains of $LiCuVO_4$ interconnected by $VO_4$ tetrahedra, where grey circle = Cu, cyan circle = V, and white circle = O. The intrachain spin exchange paths $J_{nn}$ and $J_{nnn}$ as well as the interchain spin exchange path $J_a$ are indicated by the legends "nn", "nnn" and "a", respectively.

Figure 14.    (a) The $x^2$-$y^2$ magnetic orbital of a $CuO_4$ square plane. (b) The Cu-O-Cu spin exchange interaction between nearest-neighbor $CuO_4$ square planes in a $CuO_2$ ribbon chain. (c) The Cu-O…O-Cu spin exchange interaction between next-nearest-neighbor $CuO_4$ square planes in a $CuO_2$ ribbon chain.

Figure 15.    The through-space (TS) and the through-bond (TB) interactions between the two $x^2$-$y^2$ magnetic orbitals in the Cu-O…$V^{5+}$…O-Cu interchain spin exchange $J_a$ in $LiCuVO_4$: (a) The energy split between $\psi_+$ and $\psi_-$ due to the TS interaction. (b) The bonding interaction of the V $d_\pi$ orbital with the O 2p tails of $\psi_-$ in the O…$V^{5+}$…O bridge. (c) The energy split between $\psi_+$ and $\psi_-$ due to the through-space (TS) and through-bond (TB) interactions.

Figure 16.    Three interactions controlling the strength of a DM interaction.

Figure 17.    Polar angles $\theta$ and $\phi$ defining the preferred orientation of the spin (i.e., the z′-axis) with respect to the (x, y, z) coordinate used to describe the orbital.

Figure 18.    (a) A schematic view of an isolated $CoMnO_6$ chain of $Ca_3CoMnO_6$, which is made up of the $CoO_6$ trigonal prisms containing high-spin $Co^{2+}$ ($d^7$, S = 3/2) ions and the $MnO_6$ octahedra containing high-spin $Mn^{4+}$ ($d^3$, S = 3/2) ions. (b) The occupancy of the down-spin $d$-states for a high-spin $Co^{2+}$ ion in an isolated $CoO_6$ trigonal prism.

Figure 19.    The down-spin electron configuration of a high-spin $Fe^{2+}$ ($d^6$, S = 2) at a linear coordination site that induces uniaxial magnetism.

Figure 20.    The down-spin electron configurations of a high-spin $Fe^{2+}$ ($d^6$, S = 2) at an octahedral site that induce (a) uniaxial magnetism and (b) no uniaxial magnetism.

Figure 21.    The high-spin configuration of a $Mn^{3+}$ ($d^4$) ion in an axially-elongated $MnO_6$ octahedron with the z-axis taken along the elongated Mn-O bonds.

Figure 22.    The down-spin electron configuration of a $Ni^{2+}$ ($d^8$, S = 1) ion at a trigonal prism site.

Figure 23.    The down-spin states of the low-spin $Ir^{4+}$ (S = 1/2, $d^5$) ion in (a) the axially-elongated $IrO_6$ octahedron along the 4-fold rotational axis in $Sr_2IrO_4$ and (b) the axially-compressed $IrO_6$ octahedron along the 3-fold rotational axis in $Na_2IrO_3$.

Figure 24.        (a) The structure and the down-spin $d$-states of a $CuCl_2(OH_2)_2$ complex: blue circle = Cu, green circle = Cl, medium white circle = O, and small white circle = H. (b) The down-spin electron configuration of a $Cu^{2+}$ ($d^9$, S = 1/2) ion.

Figure 25.        (a) The $CuL_2$ ribbon chain made up of edge-sharing $CuL_4$ square planes. The contributions of the metal $d$- and the ligand $p$-orbitals in the (b) yz, (c) xy and (d) $x^2–y^2$ states of a $CuL_4$ square plane.

Figure 26.        The down-spin electron configurations of a low-spin $Ir^{4+}$ ($d^5$, S = 1/2) ion at an octahedral site that induce (a) uniaxial magnetism and (b) no uniaxial magnetism.

Figure 27.        (a) An axially-compressed $VO_6$ octahedron of $R_2V_2O_7$ (R = rare earth) along the local z-direction taken along a 3-fold rotational axis. (b) The split $t_{2g}$ state of a $V^{4+}$ ($d^1$, S = 1/2) ion at each $VO_6$ octahedron. (c) A tetrahedral cluster made up of four $VO_6$ octahedra. The local z-axes of the four $VO_6$ octahedra are all pointed to the center of the $V_4$ tetrahedron. (d) The pyrochlore lattice of the $V^{4+}$ ions in $R_2V_2O_7$.

Figure 28.        (a) A view of an isolated $Sr_2IrO_4$ layer made up of corner-sharing axially-elongated $IrO_6$ octahedra approximately along the c-direction. (b) The interaction between adjacent xz orbitals (or adjacent yz orbitals) through the O $2p$ orbitals through each bent $Ir-O_{eq}-O$ bridge. (c) The interaction between adjacent xy orbitals

through the O $2p$ orbitals through each bent Ir-$O_{eq}$-O bridge. (d) The split $d$-states of a dimer made up of two adjacent $Ir^{4+}$ ions after incorporating the effect of the intersite interactions for the cases of the axially-elongated $IrO_6$ octahedra. (e) The PDOS plots for the $d$-states of $Ir^{4+}$ in $Sr_2IrO_4$ in case when the $IrO_6$ octahedra are axially elongated, where the legends (2, -2), (1, -1), and 0 indicate the sets of orbitals (xy, $x^2$-$y^2$), (xz, yz) and $3z^2$-$r^2$, respectively.

Figure 29.      (a) A projection view of a $NaIrO_3$ honeycomb layer made up of edge-sharing $IrO_6$ octahedra with Na (light blue circle) at the center of each $Ir_6$ hexagon. (b) A perspective view of how the honeycomb $NaIrO_3$ layers repeat along the c-direction in $Na_2IrO_3$, where the layer of Na atoms lying in between the $NaIrO_3$ honeycomb layers is not shown for simplicity. (c) The split $d$-states of a dimer made up of two adjacent $Ir^{4+}$ ions after incorporating the effect of the inter-site interactions.

Figure 30.      (a, b) Interactions between the Ir and Ni $3z^2$-$r^2$ states in each $NiIrO_6$ chain of $Sr_3NiIrO_6$ when the spins of adjacent $Ir^{4+}$ and $Ni^{2+}$ ions have a FM coupling in (a), and an AFM coupling in (b). (c, d) The PDOS plots for the down-spin d-states of $Ir^{4+}$ in $Sr_3NiIrO_6$ in cases when adjacent $Ir^{4+}$ and $Ni^{2+}$ ions in each $NiIrO_6$ chain have a FM coupling in (c), and an AFM coupling in (d). The legends (2, -2), (1, -1) and 0 refer respectively to  the (xy, $x^2$-$y^2$), (xz, yz) and $3z^2$-$r^2$ sets.

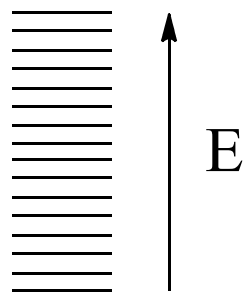

Fig. 1

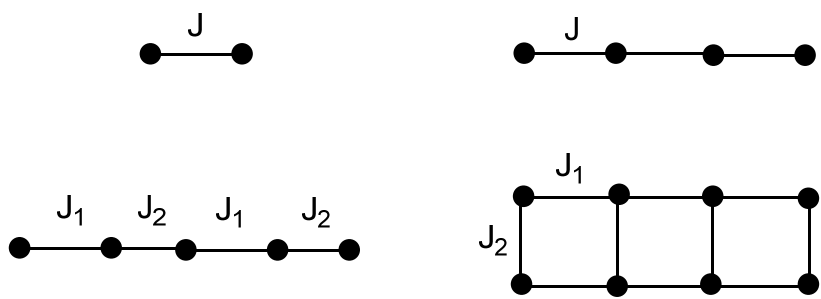

Fig. 2

(a)

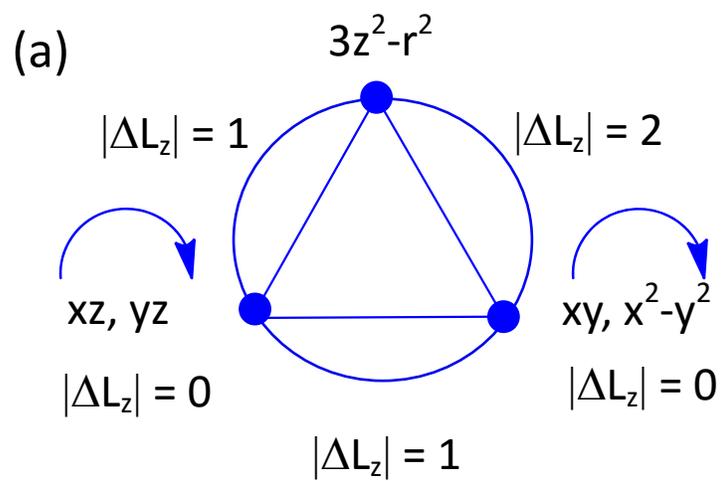

(b)

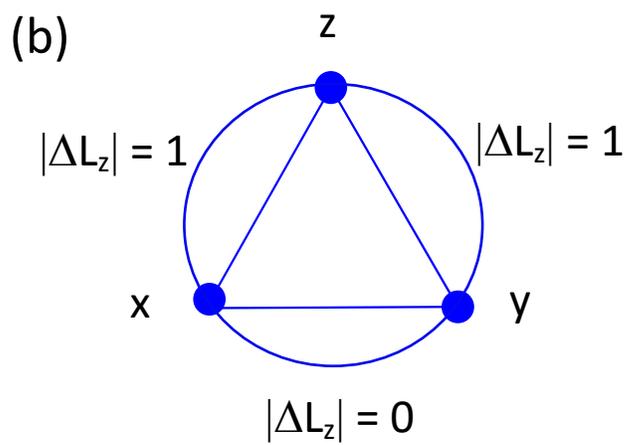

Fig. 3

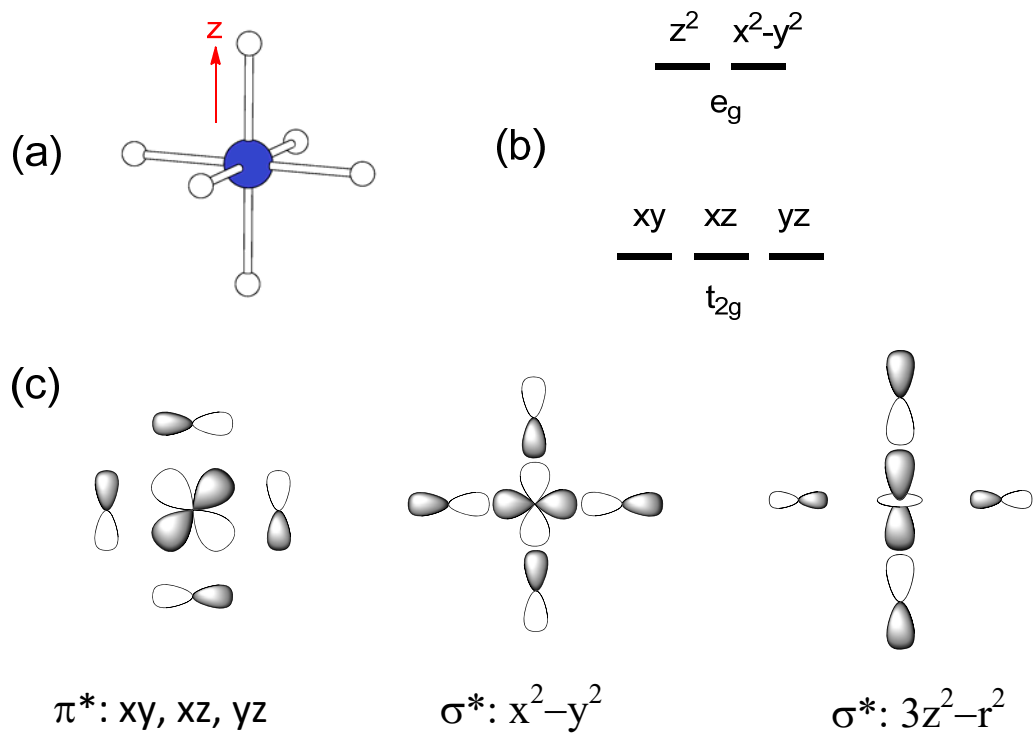

$\pi^*$: xy, xz, yz          $\sigma^*$: $x^2-y^2$          $\sigma^*$: $3z^2-r^2$

Fig. 4

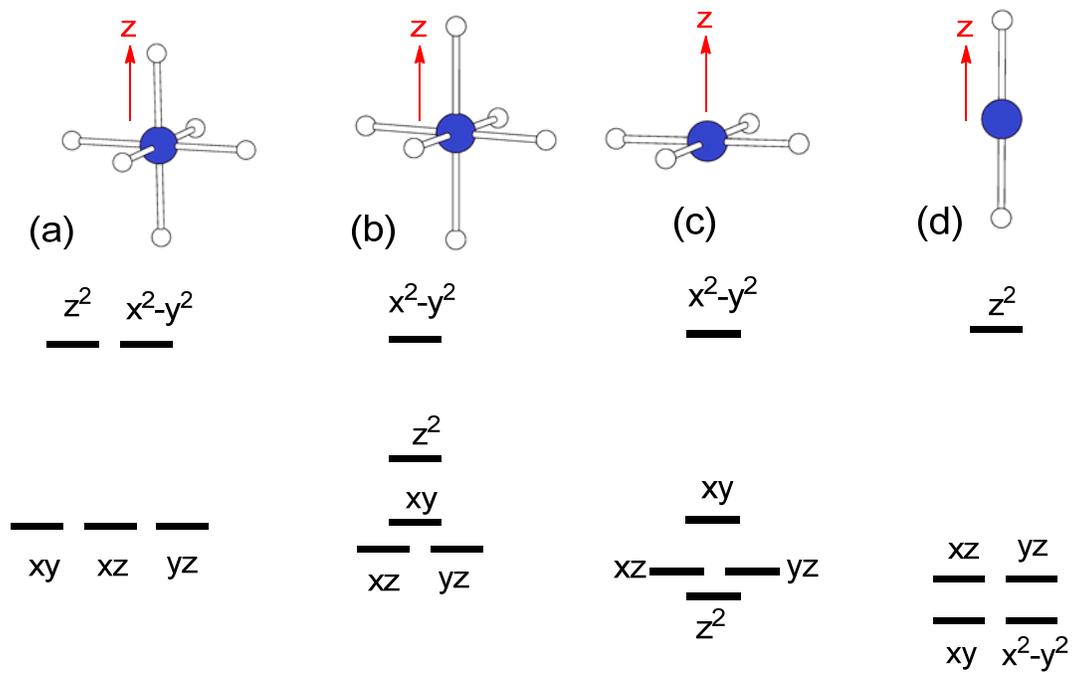

Fig. 5

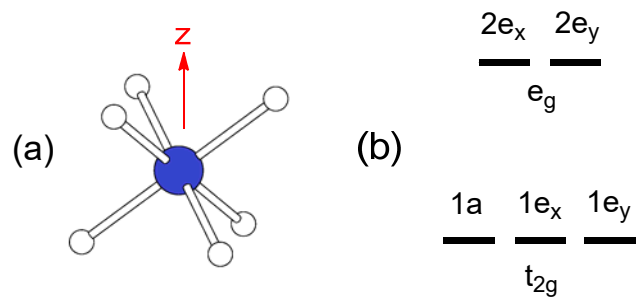

Fig. 6.

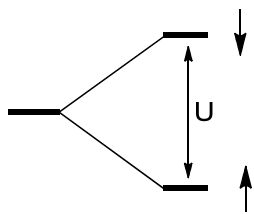

Fig. 7

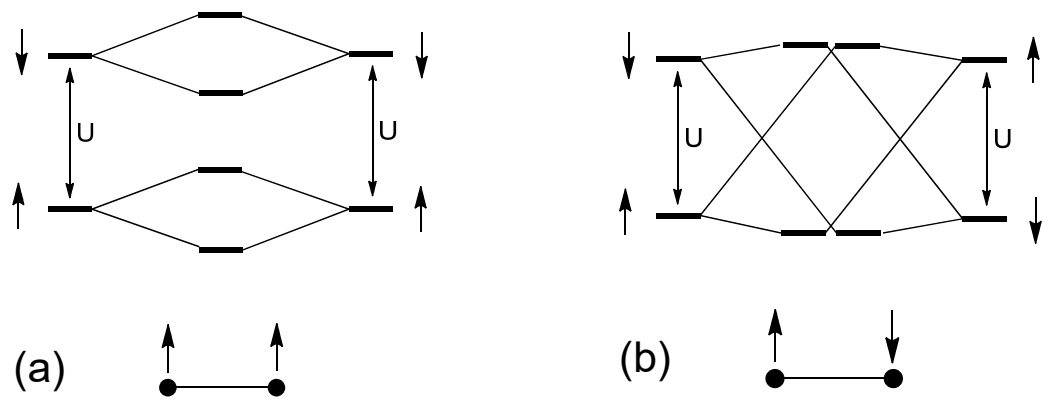

Fig. 8.

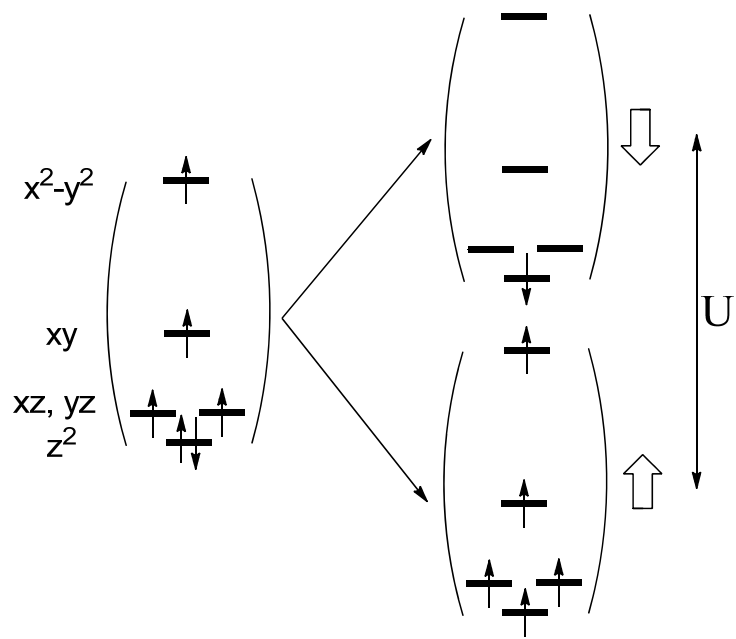

Fig. 9.

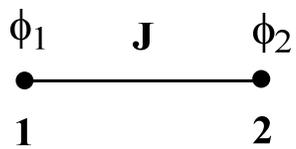

Fig. 10

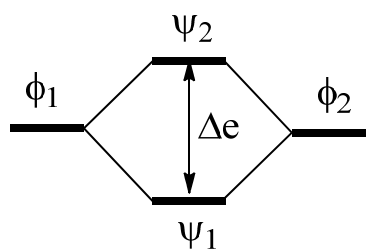

Fig. 11

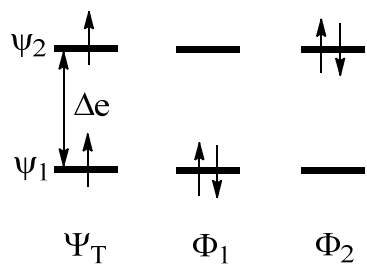

Fig. 12

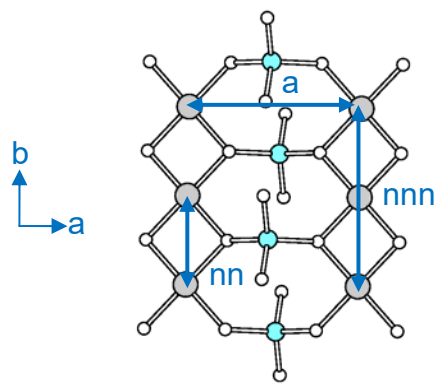

Fig. 13

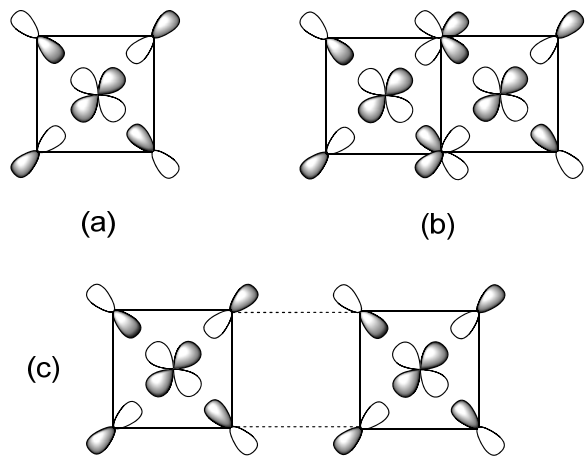

Fig. 14

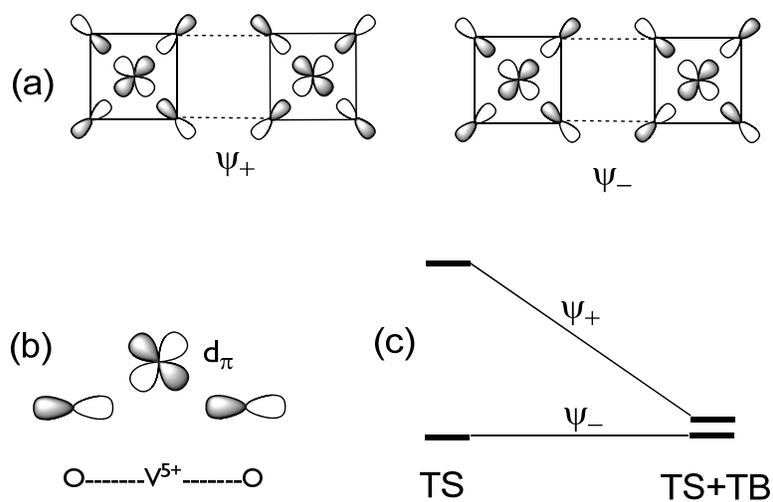

(a)   $\psi_+$   $\psi_-$

(b)   $d_\pi$

(c)   $\psi_+$   $\psi_-$   TS   TS+TB

Fig. 15

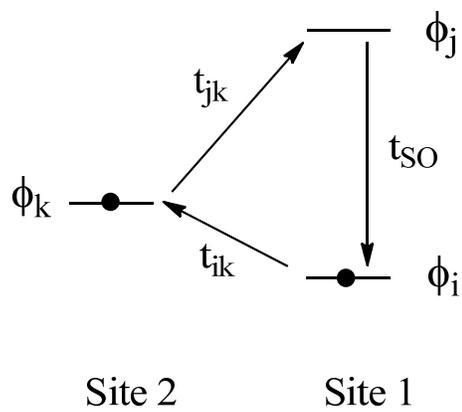

$\phi_j$

$t_{jk}$

$t_{SO}$

$\phi_k$

$t_{ik}$

$\phi_i$

Site 2   Site 1

Fig. 16

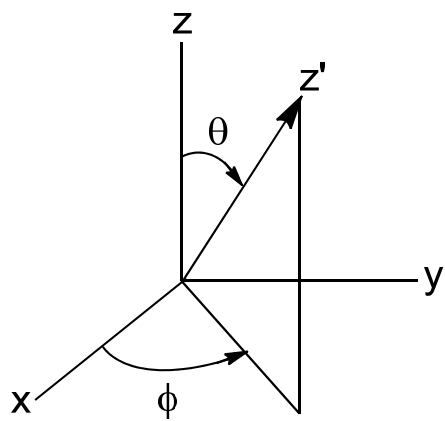

Fig. 17

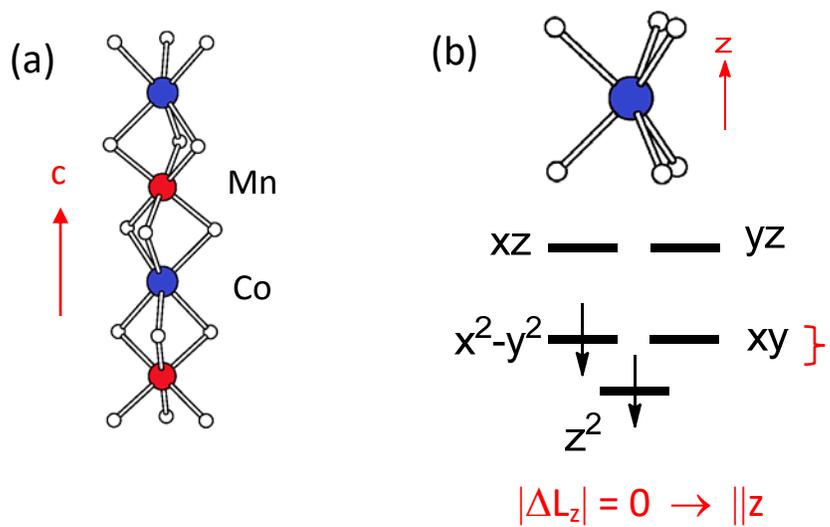

Figure 18

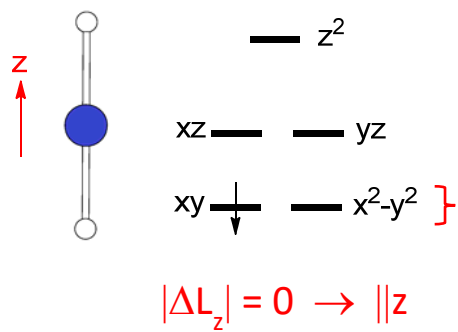

$$|\Delta L_z| = 0 \rightarrow \parallel z$$

Fig. 19

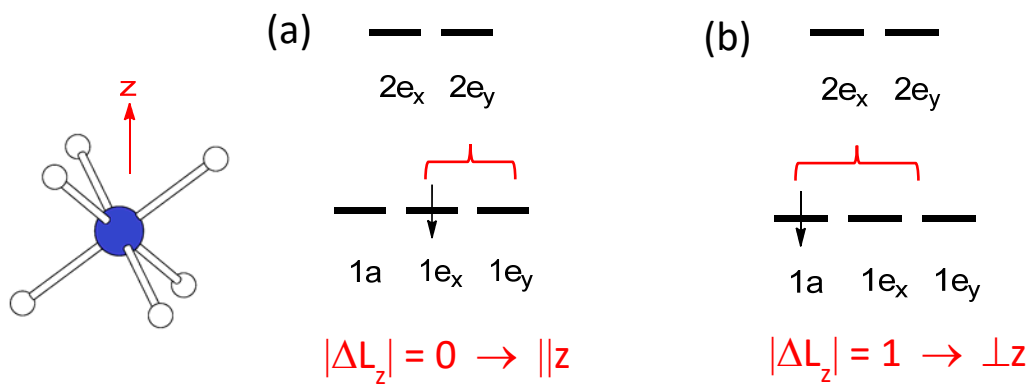

(a)

$$|\Delta L_z| = 0 \rightarrow \parallel z$$

(b)

$$|\Delta L_z| = 1 \rightarrow \perp z$$

Fig. 20

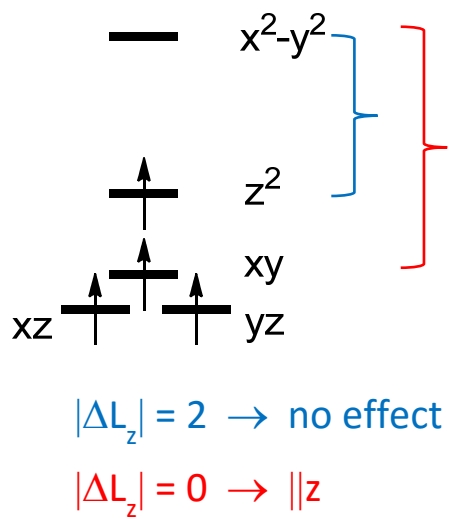

$|\Delta L_z| = 2 \rightarrow$ no effect

$|\Delta L_z| = 0 \rightarrow \parallel z$

Fig. 21

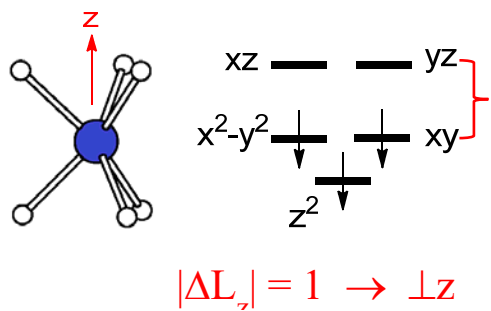

$|\Delta L_z| = 1 \rightarrow \perp z$

Fig. 22

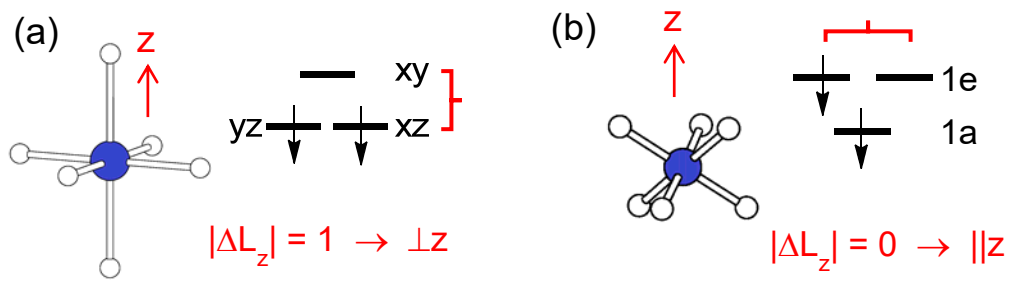

(a)    $|\Delta L_z| = 1 \rightarrow \perp z$

(b)    $|\Delta L_z| = 0 \rightarrow \parallel z$

Fig. 23

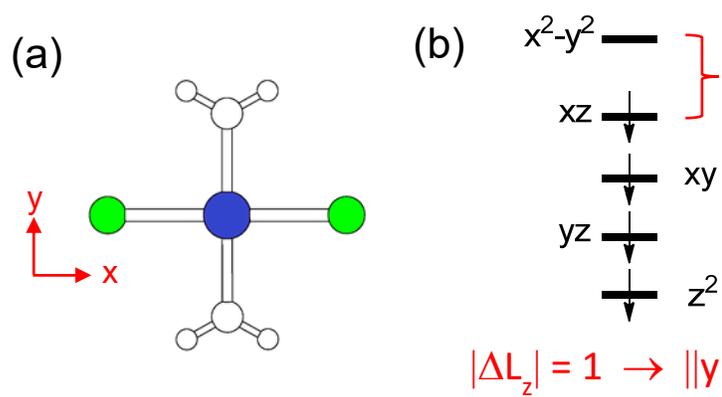

(a)

(b)

$x^2-y^2$

$xz$

$xy$

$yz$

$z^2$

$|\Delta L_z| = 1 \rightarrow \parallel y$

Fig. 24

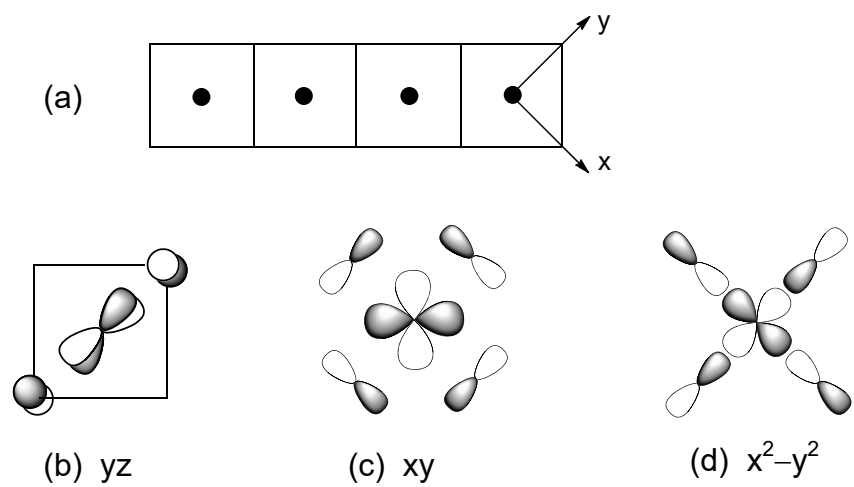

(b) yz    (c) xy    (d) $x^2-y^2$

Fig. 25

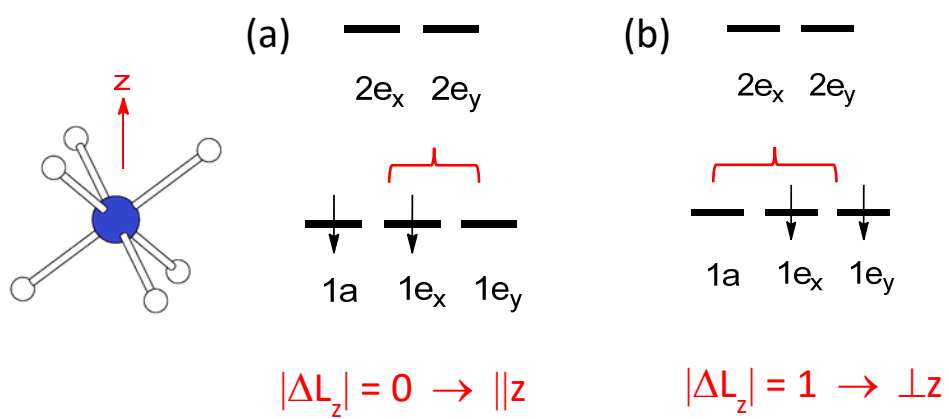

(a)

$2e_x$    $2e_y$

$1a$    $1e_x$    $1e_y$

$|\Delta L_z| = 0 \rightarrow \parallel z$

(b)

$2e_x$    $2e_y$

$1a$    $1e_x$    $1e_y$

$|\Delta L_z| = 1 \rightarrow \perp z$

Fig. 26

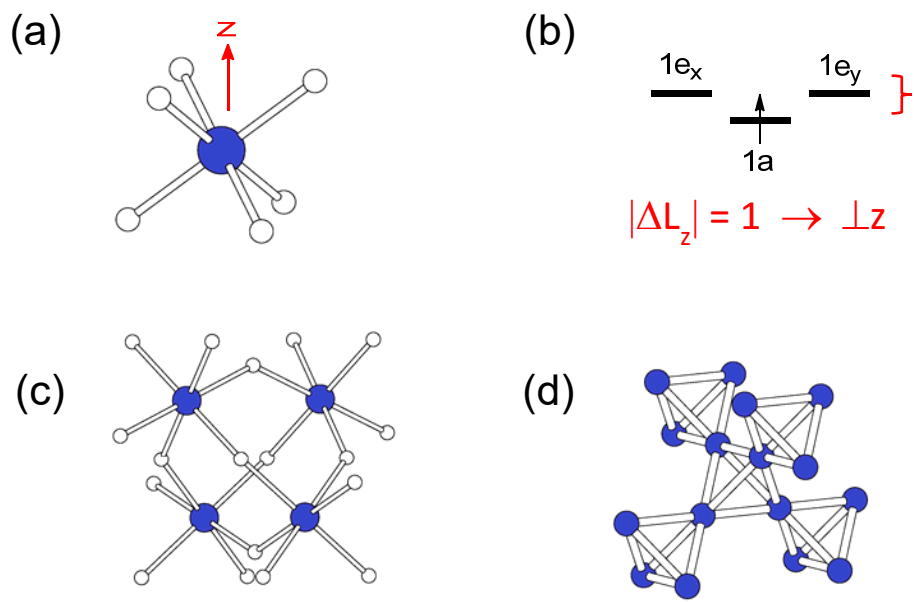

(a)

(b)

$1e_x$    $1e_y$

$1a$

$|\Delta L_z| = 1 \rightarrow \perp z$

(c)

(d)

Fig. 27

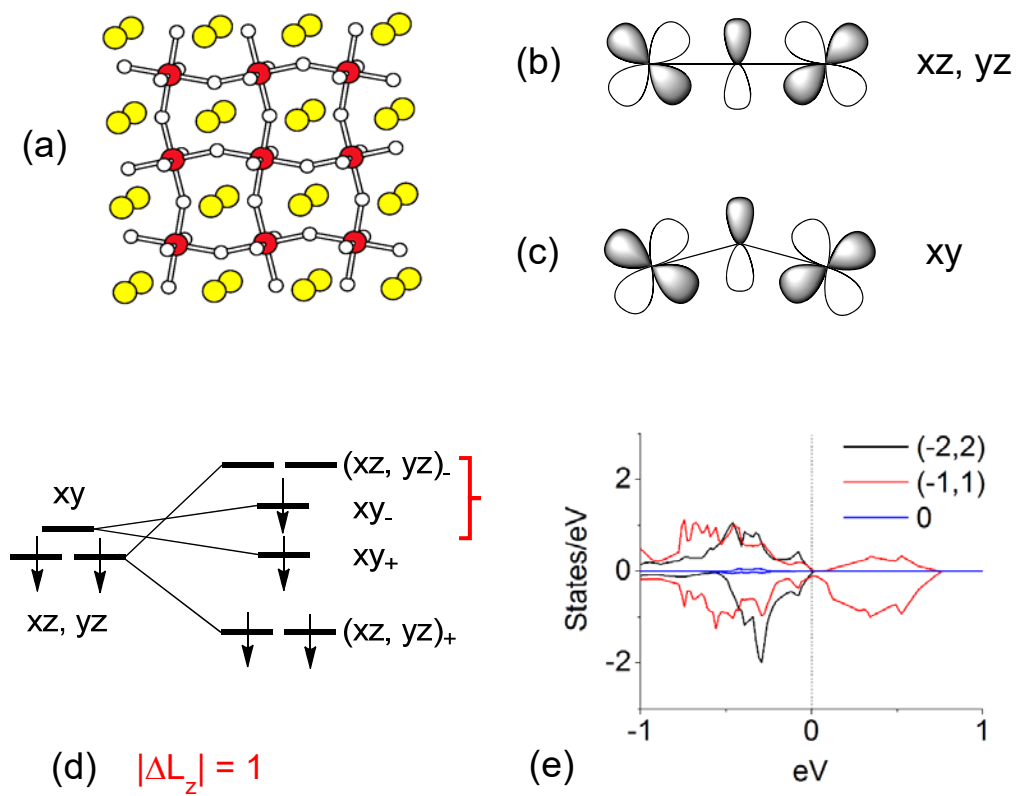

(a)

(b) xz, yz

(c) xy

(d) $|\Delta L_z| = 1$

(e)

Fig. 28

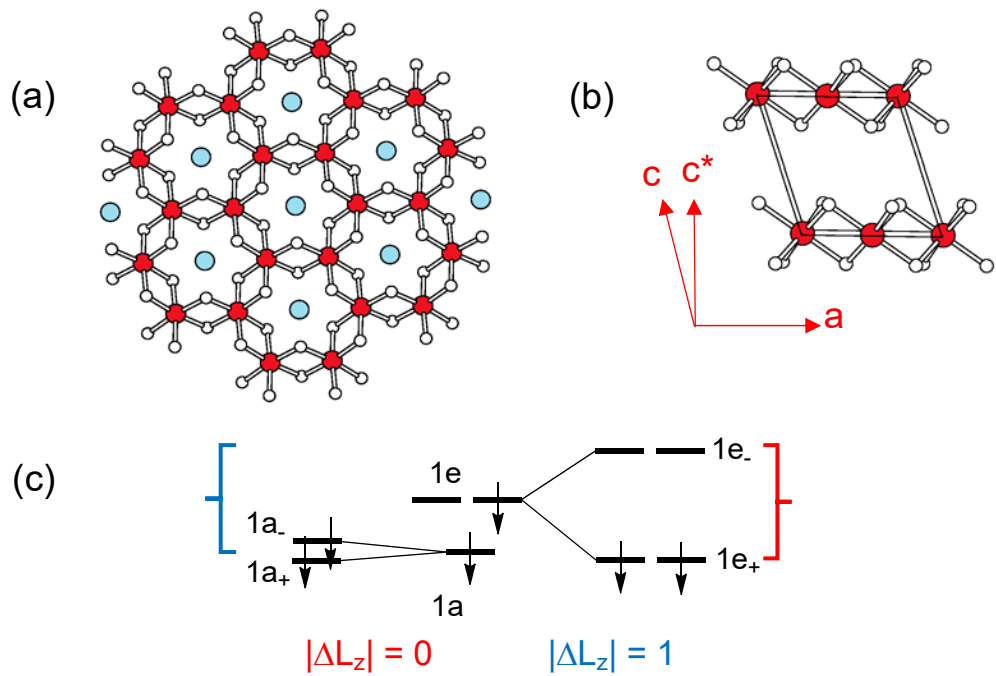

(a)

(b)

(c)

$|\Delta L_z| = 0$     $|\Delta L_z| = 1$

Fig. 29

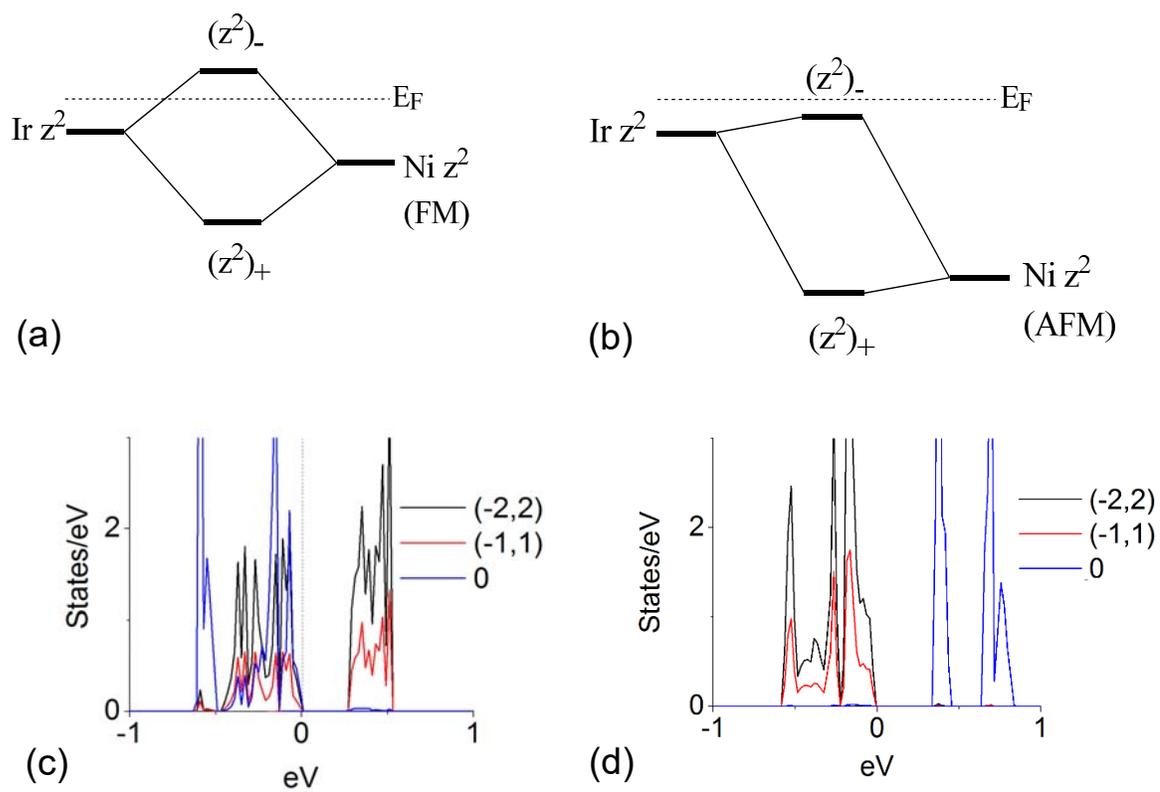

Fig. 30